# **Title:** Deep learning reveals a stronger fossil fuel influence than biomass burning in shaping remote tropospheric ozone

**Authors:** Chaoqun Ma[1], Hang Su[2*], Yafang Cheng[1*]

**Affiliations:**

[1]Aerosol Chemistry Department, Max Planck Institute for Chemistry; Mainz, 55128, Germany

[2]State Key Laboratory of Atmospheric Environment and Extreme Meteorology, Institute of Atmospheric Physics, Chinese Academy of Sciences; Beijing, China

***Email:** H.S. (suhang@mail.iap.ac.cn) or Y.C. (yafang.cheng@mpic.de)

**Abstract**: Tropospheric ozone ($O_3$) is a key greenhouse gas and atmospheric oxidant, yet its sources in the remote troposphere remain strongly debated. Observation-based tracer analyses suggest that $O_3$ attributed to biomass burning is much greater than that from fossil fuel sources (by a factor of ~2-10), contradicting state-of-the-art global models. Here we show that this discrepancy primarily arises from the strong sensitivity of tracer methods to differences in tracer lifetimes, especially after extended transport to the remote regions. To resolve this discrepancy, we develop a deep learning (DL) framework that synthesizes global observations and chemical transport model simulations. The DL approach accurately infers source contributions and reveals that fossil fuel emissions contribute over three times more $O_3$ to the remote troposphere than biomass burning. Our findings underscore that phasing out fossil fuels remains the most powerful lever for mitigating remote tropospheric ozone.

**Main Text:**

## Introduction

Tropospheric ozone ($O_3$) not only plays a central role in atmospheric chemistry [1], but also serves as a climate forcer [2] and a major air pollutant [3]. Besides the intrusion from stratosphere, the major source of tropospheric $O_3$ is the photochemical production from nitrogen oxides ($NO_x$) and volatile organic compounds (VOCs), emitted from both anthropogenic and natural sources [4,5]. $O_3$ in the remote troposphere accounts for 60–70 % of the global tropospheric $O_3$ burden [6]. It is the primary chemical source of hydroxyl radicals [7], which is the most important atmospheric detergent, sets the planet's oxidizing capacity and controls the lifetime of methane, carbon monoxide (CO) and other trace gases [8,9]. It also establishes the baseline for $O_3$ pollution in urban areas and can undermine the efforts of local emission control by elevating the background $O_3$ [10,11].

Despite its importance, the major source of $O_3$ in remote troposphere is still not well understood, particularly given the recent debate on the relative contribution of biomass burning (BB) and fossil fuel (FF) combustion to $O_3$ formation. Simulations from global chemical transport models (CTMs) generally indicate that FF sources contribute more to tropospheric $O_3$ than BB [12-16] (*SI Appendix*, Sect. S1). However, Bourgeois, et al. [17], using source-specific tracers from the ATom global aircraft observations [18], reached a different conclusion, suggesting a biomass-burning contribution approximately 2-10 times larger than that from FF sources (referred to as "*urban*" sources in [17]) and thereby posing challenges to the state-of-the-art global modelling results.

To address this debate and improve understanding of $O_3$ formation in remote regions, we first evaluated the underlying assumptions and performance of the observation-based tracer approach of Bourgeois, et al. [17] (OT method hereafter) in quantifying source-$O_3$ relationships (i.e., the contribution of different sources to $O_3$ formation). To overcome the limitations identified in the OT methods, we developed a new deep learning (DL) framework that integrates information from both global CTM simulations and observations. The new method accurately predicts the $O_3$ contribution from BB ($O_{3,BB}$) and FF ($O_{3,FF}$) sources and substantially outperforms the OT method. Applied globally using observed tracer mixing ratios, the DL framework reveals a consistently larger influence of fossil fuel sources than biomass burning in shaping remote tropospheric $O_3$ across most regions and seasons worldwide.

**Tracer lifetime difference explains discrepancies in $O_3$ attribution**

To evaluate the performance of the OT method from Bourgeois, et al. [17], we focus on the tropical and southern extratropical Atlantic region during December 2017–January 2018. This period features two geographically separated BB source regions with different emission intensities, enabling a clear comparison of their associated $O_3$ formation. As shown in Fig. 1A, BB emissions during this period are much stronger in northern Africa and weaker in South America. Given the prevailing transport patterns indicated by the wind fields (Fig. 1A), $O_{3,BB}$ is thus expected to be higher over the tropical Atlantic (downwind of northern African fires) than over the extratropical Atlantic (downwind of South American fires). This expected pattern is indeed well captured by the CTM-simulated $O_{3,BB}$ obtained by differencing simulations with and without BB emissions (Fig. 1A; the CTM used in this study is GEOS-Chem v13.3.4, see *Methods*). However, when the OT method is applied to the simulated tracer concentration, it yields an opposite pattern with an anomalous high $O_{3,BB}$ belt over the extratropical Atlantic and low $O_{3,BB}$ over the tropical Atlantic (Fig. 1B). This counter-intuitive behavior of the OT method also persists under other scenarios (e.g., *SI Appendix*, Fig. S1), indicating an intrinsic limitation of the method itself.

To pinpoint the cause, we revisited the OT method and examined its underlying assumptions. In the OT method, $O_{3,BB}$ in an air mass is inferred as $\Delta O_3 \times (ER \times \Delta HCN)/\Delta CO$, where $\Delta CO$ ($\Delta HCN$, $\Delta O_3$) denotes the difference between the mixing ratio of CO (HCN, $O_3$) and its background level. HCN (hydrogen cyanide) serves as a BB-specific tracer, and ER is the emission ratio of CO to HCN in the BB emission inventory. Assuming similar lifetimes of HCN and CO, $(ER \times \Delta HCN)/\Delta CO$ approximates the fraction of CO originating from BB and, by extension, the fractional BB influence on an air mass [17]. In reality, however, CO has a shorter lifetime and decays more rapidly than HCN in the atmosphere [19,20]. This lifetime difference artificially inflates $\Delta HCN/\Delta CO$ and therefore leads to an overestimation of the BB contribution. The effect is particularly pronounced in regions far from the source after extended transport, where the influence of lifetime difference is magnified. Over the extratropical Atlantic, for instance, the inferred apparent transit time is ~80–180 days (*SI Appendix*, Sect. S2), nearly an order of magnitude longer than the 2–28 days estimated from trajectory analyses (17). This discrepancy most likely reflects substantial mixing with aged air masses along the trajectory and/or from earlier transport episodes not captured by the trajectories (*SI Appendix*, Sect. S2). After applying a first-order, proof-of-principle correction within the OT method to account for the CO-HCN lifetime difference, implemented by adjusting the HCN mixing ratio using an exponential-decay function and an assumed initial concentration of HCN (*SI Appendix*, Sect. S2), the inferred $O_{3,BB}$ distribution (Fig. 1C) becomes more consistent with expectations, exhibiting spatial patterns similar to those from the CTM-based attribution (Fig. 1A). This simplified correction is intended to provide an intuitive demonstration of how tracer lifetime differences affect the inferred attribution. Building on this insight, the following section introduces a more rigorous and physically and chemically grounded approach that explicitly accounts for tracer lifetime effects using a deep learning framework.

## Deep learning framework for accurate $O_3$ source attribution

To overcome the limitations identified in the OT approach and to resolve the discrepancy between observation- and model-based estimates, we developed a new deep-learning (DL) framework that integrates global CTM simulations with observations and directly learns the relationship between source-specific $O_3$ contributions and tracer mixing ratios. (Fig. 2A). The DL model is trained on global CTM simulations with full chemistry and transportation under an ensemble of emission scenarios (one baseline and three perturbed scenarios in which BB and FF emissions are varied by a factor of ten and two, respectively, to account for potential uncertainties in emissions; see *Methods*). This enables the DL model to capture the tracer-source relationships and to implicitly account for the nonlinear $O_3$-precursor relationships, both governed by chemistry and transport processes across a wide range of emission conditions, thereby reducing its dependence on any specific emission scenarios. By applying observed tracer data as input to the trained DL model, we obtained a model-informed and observation-constrained estimate of the contributions of FF and BB emission to remote tropospheric $O_3$.

To assess the performance of the DL model, we conducted cross-validation against held-out data. As shown in Figs. 2B and 2C, the narrow spread and high density of points near the 1:1 line indicate strong agreement between the DL-predicted $O_3$ relative contribution and the ground truth. The DL model achieves coefficients of determination ($R^2$) exceeding 0.93 and mean relative bias (MRB) close to zero. In contrast, $O_3$ contributions predicted by the OT method [17] exhibit substantial larger scatter (Figs. 2D and 2E), with $R^2$ reduced to 0.69 for BB contributions and 0.77 for FF contributions. It also shows pronounced positive bias, with MRB values of 4.1% for FF sources and 56.0% for BB sources. Here, the bias in BB contribution is larger than that in FF because tetrachloroethylene ($C_2Cl_4$), the tracer for FF sources (*SI Appendix*, Sect. S2), has a lifetime more similar to CO than HCN. This imbalance in biases causes the OT method to spuriously inflate the relative importance of BB compared with FF in contributing to $O_3$ (Figs. 2D and 1B). Unlike the OT method, the DL model captures the elevated $O_{3,BB}$ over the tropical Atlantic without introducing a spurious overestimation (Fig. 1A; *SI Appendix*, Fig. S2). Overall, the new DL framework delivers higher accuracy, lower bias, and improved spatial consistency than the OT method, establishing it as a more reliable tool to fuse observations for $O_3$ source attribution.

## Fossil fuel matters more than biomass burning in shaping remote tropospheric ozone

Applying the DL model to ATom aircraft observations [18] reveals distinct spatial and seasonal distribution in source-specific $O_3$ contributions in remote troposphere globally. As shown in Fig. 3, $O_{3,BB}$ is generally higher near major fire regions, e.g., west of North America in Northern Hemisphere (NH) summer, east of Australia in Southern Hemisphere (SH) summer and west of tropical Africa in both seasons, closely tracking seasonal and spatial variation of wildfire activity. In contrast, $O_{3,FF}$ is consistently higher in the NH than in the SH, reflecting the dominance of anthropogenic emissions in the NH.

Figure 4 further compares contributions of FF and BB to remote tropospheric $O_3$ averaged across different regions and seasons. As shown in Fig. 4A, the BB contribution inferred by the DL model is markedly lower than that calculated using the conventional OT method [17], yet it aligns more closely with the CTM-based source attribution. Notably, without correcting for tracer lifetime differences, the OT method infers a relatively large negative FF contribution in Tropical Atlantic.

The DL model, however, overcomes this limitation of the OT method and yields physically plausible $O_3$ contributions across all regions (Fig. 3).

As shown in Fig. 4B, in the tropical Pacific and the NH extratropics throughout the year, FF contributes more than twice as much $O_3$ as BB. FF also exerts a greater influence in the SH extratropics during SH fall and winter. The contrast is starkest in the NH extratropics during NH fall and winter (9.3 ppb of $O_{3,FF}$ vs. 0.9 ppb of $O_{3,BB}$), consistent with strong anthropogenic consumption of fossil fuel and minimal BB activities in the NH extratropics during winter time [21,22]. Even over the tropical Atlantic, often considered heavily impacted by BB smoke [23-25], FF contribution is still slightly more than BB. Only in the SH extratropics during austral spring and summer does BB exceed FF, and even there only marginally.

For the global average contribution, it is important to note that the ATom measurements did not uniformly sample the remote troposphere. For example, the western Pacific and offshore of Angola, areas with relatively high contributions from FF [11,26] and BB [27] emissions, respectively, were not covered. As a result, the ATom campaign average is very likely to underestimate the global contributions of both FF and BB sources to the remote troposphere. This interpretation is consistent with the model simulations shown in *SI Appendix*, Fig. S3, where the global mean $O_3$ contributions across the entire remote troposphere are 16% (FF) and 8% (BB) higher than the averages obtained by sampling only alone the ATom flight paths. We therefore adjusted the ATom-based average using these simulated differences between global remote troposphere and ATom-sampled coverage. After this correction, we found that $O_3$ contribution from FF emissions exceeds those from BB sources globally by more than a factor of three (Fig. 4B).

**Discussion and implications**

Our analysis demonstrates that the discrepancies in $O_3$ attribution between the OT method and the CTM-based approach mainly originate from the strong sensitivity of tracer methods to differences in tracer lifetimes. We evaluated the potential impact of uncertainties in the CTM related to the uptake of hydroperoxyl on aerosols and chemistry of nitrous acid, terpenes, furans, and other VOC species, as suggested by Bourgeois, et al. [17] (*SI Appendix*, Sect. S3). As shown in *SI Appendix*, Fig. S4, the combined influence of these factors is approximately eight times smaller than the tracer lifetime issue inherent to the OT method and is therefore insufficient to explain the large discrepancies in $O_3$ attribution between the OT method and CTM simulations.

By leveraging DL model with CTM simulations and ATom observations, we show that fossil fuel emissions contribute over three times more $O_3$ to the remote troposphere than biomass burning. Importantly, the final $O_3$ source attribution uses observed tracer concentrations as input to the trained DL model, providing an observational constraint on the inferred contributions. Our finding underscores that phasing out fossil fuels remains the most effective strategy for mitigating remote tropospheric $O_3$ under present-day conditions. In the longer term, however, the reduction of fossil fuel use and increasing frequency and intensity of wildfires under global warming [28,29] may elevate the impact and importance of biomass burning sources.

The new DL framework represents a significant advancement over the traditional observation-based tracer methods for global $O_3$ source apportionment. Conventional OT approaches suffer from several limitations that undermine their accuracy, including differences in lifetimes between tracers and the targeted species, tracer emissions from unintended sources, inconsistencies in assumed background concentrations, and uncertainties in tracer-target relationships arising from complex atmospheric processes. The DL framework overcomes these issues by leveraging ensemble CTM simulations to capture complex chemistry, atmospheric transport, and varying emission conditions, thereby avoiding assumptions about tracer lifetimes and background levels. Moreover, its application is not restricted to tracer observations such as HCN and $C_2Cl_4$. It can be readily extended to additional tracers, including levoglucosan for BB [30], rare earth elements for motor vehicle emissions [31], carbonyl sulfide for biogenic organics [32], calcium for mineral dust [33] or isotopic signatures for multiple source types [34-36]. Taken together, our framework demonstrates the potential of deep learning methods in advancing atmospheric environment research by enabling accurate global source apportionment and improved capability for diagnosing and managing air pollutant and greenhouse gases.

## Figures and Tables

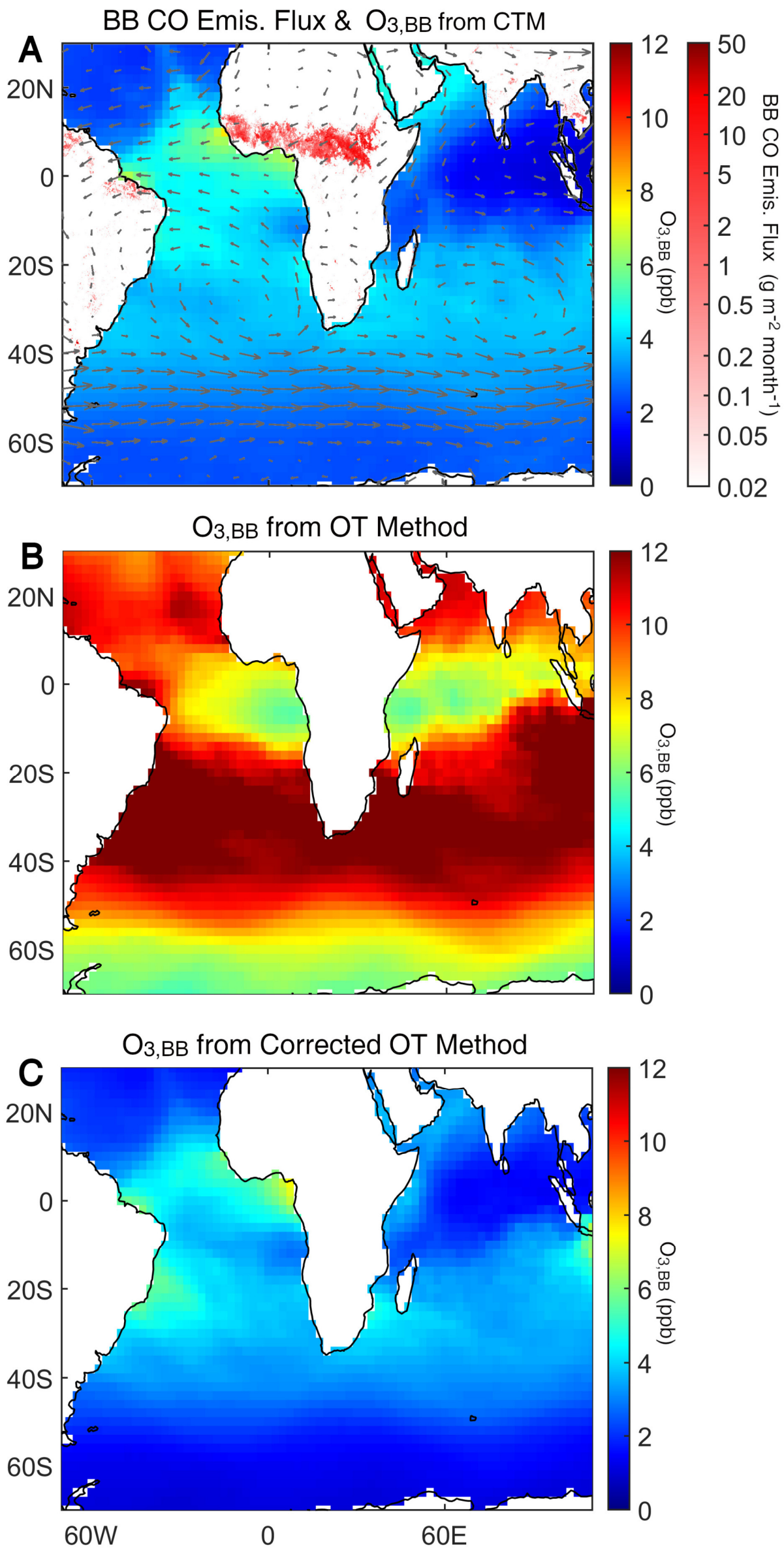

**Figure 1. Performance of the observation-based tracer method in source attribution of remote tropospheric ozone.** (**A**) Emission flux of carbon monoxide (CO) from biomass burning (BB) and the corresponding ozone ($O_3$) contribution from BB ($O_{3,BB}$) at 500 hPa over the ocean, obtained by differencing chemical transport model (CTM) simulations with and without BB emissions from the Base scenario (see *Methods*). Both fields are averaged over December 2017 to January 2018. Overlaid wind vectors represent the mean flow from the surface to the 500 hPa level. (**B**) $O_{3,BB}$ at 500 hPa calculated using the observation-based tracer (OT) method [17] with CTM-simulated tracers as input, averaged over the same period as in (**A**). (**C**) Same as (B), but with a correction applied for accounting tracer lifetime differences.

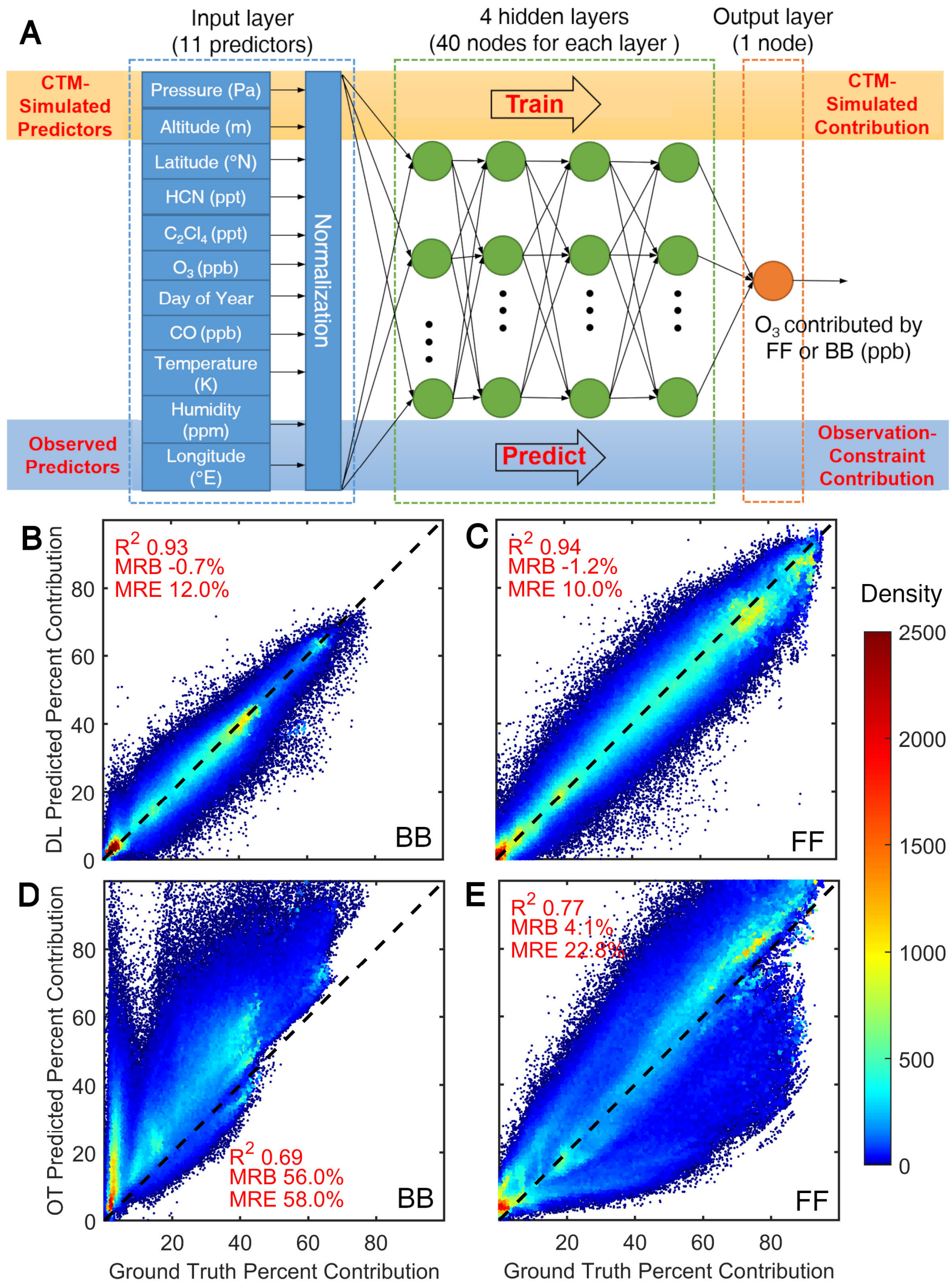

**Figure 2. Framework of the deep learning model and its performance in predicting ozone source contributions compared with the observation-based tracer method.** (**A**) Architecture of the deep learning (DL) model with its training and prediction procedure. The DL model is a multilayer perceptron network with 11 input variables and 1 output variable; the models that predict ozone ($O_3$) contributed by biomass burning (BB) and fossil fuel (FF) sources are independent of each other during training and validation. The output of the model is the absolute mixing ratio of $O_3$ contributed by BB ($O_{3,BB}$) or FF ($O_{3,FF}$) source. (**B** and **C**) Performance of the DL model in inferring $O_3$ contributions from BB (left column) and FF (right column) sources, respectively. Density scatterplots show predicted versus ground true percent $O_3$ contributions from a grid-based 10-fold cross-validation. (**D** and **E**) Corresponding results from the observation-based tracer (OT) method [17]. The 1:1 line is shown as a black dashed line. The coefficient of determination ($R^2$), mean relative bias (MRB), and mean relative error (MRE) are reported for each case.

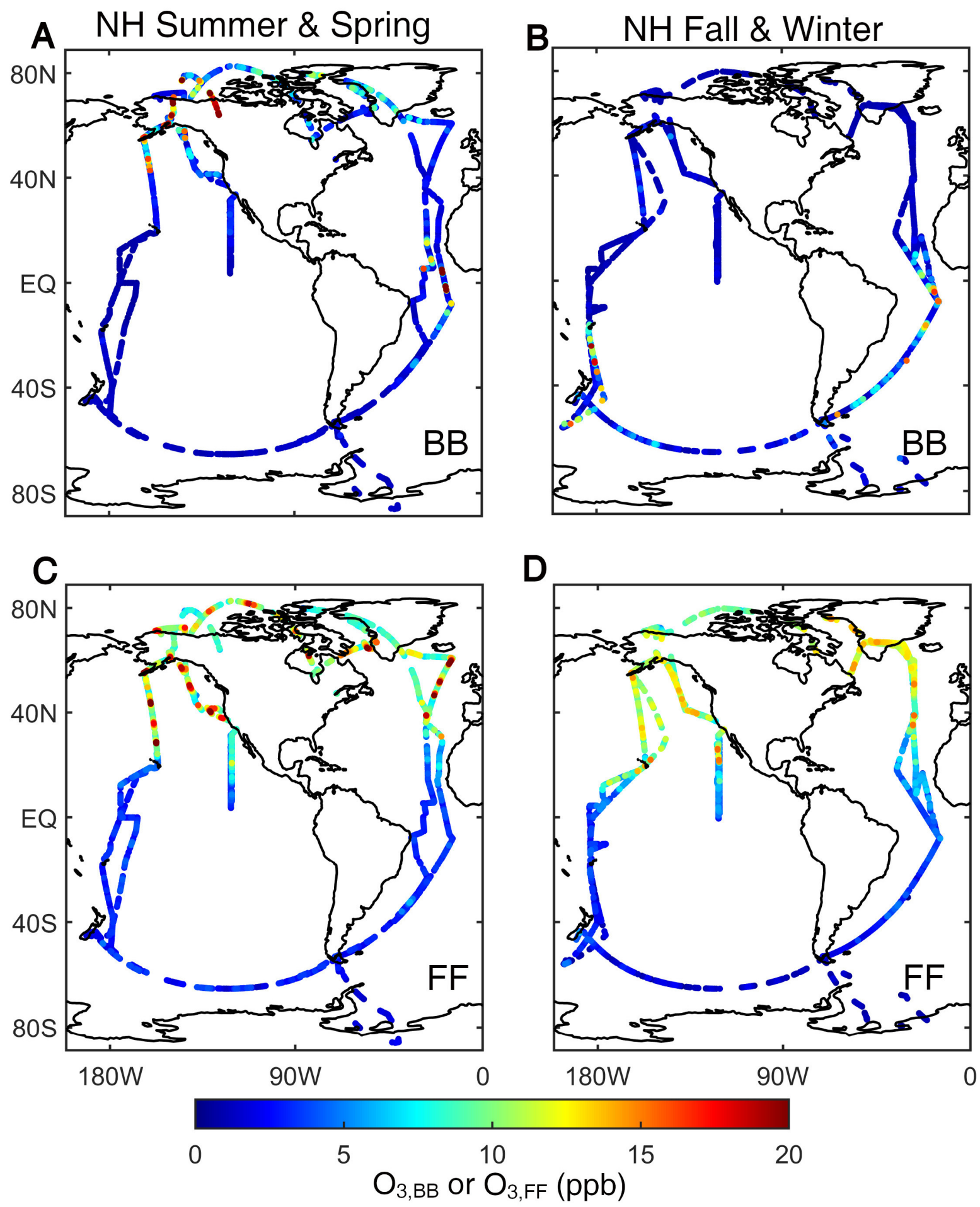

**Figure 3. Spatial and seasonal distribution of remote tropospheric ozone attributed to biomass burning and fossil fuel sources.** Colored circles indicated the absolute mixing ratio of ozone contributed by biomass burning ($O_{3,BB}$; top row) or fossil fuel ($O_{3,FF}$; bottom row) predicted by deep learning (DL) model using ATom observations as input. Data are grouped by season: Northern Hemisphere (NH) summer and spring (**A** and **C**) correspond to March-August, whereas NH fall and winter (**B**, **D**) correspond to September-February.

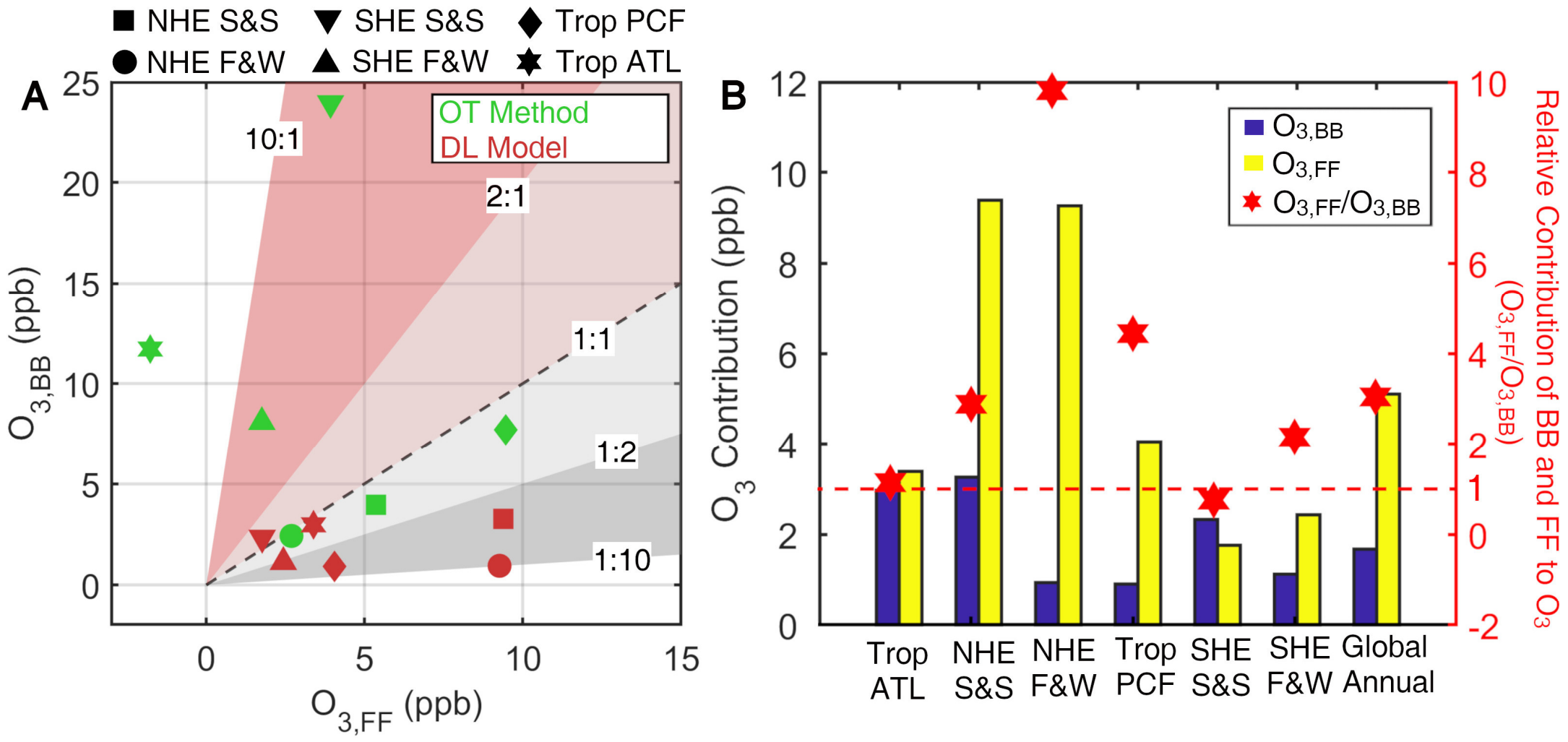

**Figure 4. Regional and global contributions of fossil fuel and biomass burning sources to remote tropospheric ozone across different seasons**. (**A**) Averaged ozone ($O_3$) contribution inferred from the observation-based tracer (OT) method [17] (green) and the deep learning (DL) model (red). Contributions from biomass burning (BB) and fossil fuel (FF) are denoted as $O_{3,BB}$ and $O_{3,FF}$, respectively. Data from the Northern Hemisphere (NH) extratropics (NHE; >20°N) and Southern Hemisphere (SH) extratropics (SHE; <20°S) are divided into spring-summer (S&S) and fall-winter (F&W) seasons. NHE S&S and SHE F&W both correspond to March-August, whereas NHE F&W and SHE S&S correspond to September-February. The tropics (Trop; 20°S-20°N) are further separated into the tropical Atlantic (ATL) and tropical Pacific (PCF) regions. (**B**) Comparison of regional and global contributions of BB and FF to remote tropospheric $O_3$. Regional and seasonal averages are based on ATom samples available within each region and season, the same as in (**A**). The global annual value is obtained by averaging all ATom samples and adjusting the result using the simulated difference between the global and ATom-sampled remote troposphere (see *SI Appendix*, Fig. S3). Red hexagrams indicate the corresponding ratios between FF and BB contributions.

## Methods:

### Model Simulations

The mixing ratio of $O_3$ and other tracers were simulated using GEOS-Chem v13.3.4 on a global grid of 2° × 2.5°, driven by the MERRA-2 (Modern-Era Retrospective analysis for Research and Applications, Version 2) meteorological fields [37]. The simulations employed the standard "full-chemistry" mechanism (i.e., NOx-Ox-hydrocarbons-aerosols-halogens mechanism, https://geos-chem.readthedocs.io/en/latest/geos-chem-shared-docs/simulations/fullchem.html), which has been widely adopted in studies of tropospheric ozone [38,39]. Anthropogenic emissions were taken from the Community Emissions Data System (CEDS) [40] and the Aviation Emissions Inventory Code (AEIC) [41], while biomass burning (BB) emissions were taken from the Quick Fire Emissions Dataset 2 (QFED2) [42]. As shown in previous studies [20,43-47], HCN and $C_2Cl_4$ are long-lived tracers predominantly emitted by BB and industrial processes, respectively. Following Bourgeois, et al. [17], we used these compounds as tracers for BB and fossil fuel (FF) sources. To represent emissions and atmospheric transport of these tracers in GEOS-Chem, we implemented $C_2Cl_4$ as a passive tracer with a lifetime of 101 days [47] and an emission ratio of 0.03 pptv ppbv$^{-1}$ per unit anthropogenic CO, and HCN as a passive tracer with a lifetime of 150 days [20] and an emission ratio of 5.7 pptv ppbv$^{-1}$ per unit CO emission from BB, following Bourgeois, et al. [17]. GEOS-Chem reproduces the ATom-observed HCN and $C_2Cl_4$ abundances with relatively small mean biases (<13%) in the remote troposphere (*SI Appendix*, Fig. S5), indicating that the model captures the large-scale variability of these tracers reasonably well.

To match the ATom campaign period, the simulation period covered January 1, 2016 and to May 31, 2018 with the first seven months as spin-up. Hourly outputs were archived every three hours. To ensure sufficient coverage of the training dataset, we constructed an ensemble of four CTM simulation scenarios:

(1) Base: This scenario used the configuration described above to represent the standard model setup. The consistency of simulated $O_{3,FF}$ ($O_{3,BB}$) distributions with observed $C_2Cl_4$ (HCN) (*SI Appendix*, Fig. S6), together with the simulated fraction of the global tropospheric $O_3$ burden over remote oceans (60–70 %, consistent with Bourgeois, et al. [6]), supports the reliability of the Base model setup in reproducing large-scale $O_3$ distributions.

(2) BB10abovePBL: To account for potential underestimation of BB emissions, this scenario scaled BB emissions by a factor of 10 and released them entirely above the planetary boundary layer (PBL). The scale factor 10 exceeds typical inventory uncertainties [48,49] but was still applied to yield a BB $O_3$ contribution comparable to that reported by Bourgeois, et al. [17] (*SI Appendix*, Fig. S4 and S7). The high scale factor also helps avoid extrapolation when applying the DL model.

(3) FF2: To address the potential underestimation of FF emissions amount, anthropogenic emissions were doubled in this scenario. The factor of two is larger than typical uncertainties of anthropogenic emissions [21,40] but again helps avoid extrapolation.

(4) FFHCN: This scenario was designed to account for potential uncertainties in tracer emission ratios. Given the slight underestimation of HCN mixing ratios in the Base simulation (*SI Appendix*, Fig. S4) and possible emission of HCN from sources collocated with fossil fuel usage [44,50], this scenario has HCN added to anthropogenic emissions as 5.7 pptv ppbv$^{-1}$ of anthropogenic residential CO emission. This led to an overestimation of HCN mixing ratios, representing an upper bound for the tracer emission ratio of HCN. More scenarios with possible HCN or $C_2Cl_4$ emission ratios were also discussed in *SI Appendix*, Sect. S4, but not included in the training dataset here because they had limited influence on the overall conclusions.

For each scenario, three simulations were conducted: (i) all emissions included, (ii) without anthropogenic emissions, and (iii) without BB emissions. CTM-simulated $O_{3,BB}$ or $O_{3,FF}$ can then be calculated as the difference between the full simulation and those with BB or anthropogenic emissions removed. For simplicity, anthropogenic fossil fuel usage (referred to here as the "FF" source) is represented by total anthropogenic emissions, since other anthropogenic activities such as biofuel use contribute only a minor fraction [51]. The impact of $O_3$ chemistry nonlinearity [12] on the results of contribution is quantified and found to be limited in *SI Appendix*, Sect. S5, which is expected given $O_3$ in remote area is currently controlled linearly by $NO_x$ availability [52]. Together, these four scenarios yield $O_3$ contribution estimates that encompass those inferred from observations (*SI Appendix*, Fig. S7), indicating a sufficient coverage of the training dataset.

**Extraction of Training Data.**

For each scenario, the model simulation produced approximately $3.3 \times 10^9$ data points (144 longitude × 91 latitude × 47 vertical levels × 668 days × 8 time-steps per day). Following the definition of remote troposphere from Bourgeois, et al. [17], we excluded data over continental grids and those influenced by marine boundary layer or stratospheric intrusions. After that, around $1.0 \times 10^9$ usable samples remained. From those samples, we randomly selected $2 \times 10^5$ samples per scenario for subsequent use. Because most $O_3$ contribution values were close to zero, a purely random selection risked underrepresenting extreme value. To address this imbalance, we applied a data augmentation strategy: bins with many samples were subsampled, while bins with few samples were oversampled by duplicating cases. In total, this yielded $2 \times 10^5 \times 4$ scenarios = $8 \times 10^5$ samples for subsequent training and validation.

**Deep Learning Model**

As shown in Fig. 2A, the DL model used in this study is a multilayer perceptron network with 11 input nodes, four hidden layers of 40 nodes each, and one output node. Neurons were activated with sigmoid function and penalized by an L2-regulization with coefficient of $1 \times 10^{-5}$. Parameters were optimized by adaptive moment estimation (adam) optimizer to minimize a loss function of mean squared error. Training was conducted over 10 epochs with a batch size of 128. While alternative architectures and hyperparameters were explored, no significant performance improvements were observed. Although the model may not be fully optimized, its configuration proved sufficient for accurate $O_3$ attribution.

We selected 11 input variables including temperature, humidity and mixing ratios of $O_3$, CO, $C_2Cl_4$, HCN along with their sampling time and position (Fig. 2A). All input variables were first normalized to zero mean and unit standard deviation using statistics derived from the training dataset. Although geographic position and time were included as inputs, model interpretation analyses (*SI Appendix*, Sect. S6) demonstrate that they have only minor impact compared with other input variables, confirming that the trained model did not simply memorize training data by location or time. The network included a single output node to predict the ozone contribution from either BB or FF sources, which means two models were developed and validated independently, one for each source.

To evaluate the model performance, a grid-based 10-fold cross-validation approach was adopted. During the cross-validation, model grids over ocean were randomly divided into 10 parts. During each fold, samples from one part of the grids was withheld to validate the network trained on the remaining nine parts. For consistency, the normalization of input variables was applied with statistics derived from their corresponding training datasets. The overall result is shown in Figs. 2B and 2C as density scatter plots with corresponding $R^2$, mean relative error (MRE) and mean relative bias (MRB). *SI Appendix*, Figs. S8 and S9 also provide the results from other approaches of validation.

**Observation Data.** The NASA Atmospheric Tomography (ATom) mission provides in situ observations of chemical, meteorological, and aerosol microphysical properties measured by extensive instruments deployed on the NASA DC-8 aircraft [18]. The data covers the period from August 2016 to May 2018 and a vertical range of 0.2–13 km in altitude. In this study, we used the two-minute average measurements of the corresponding 11 input variables for our trained DL model to predict the individual contributions of FF and BB emissions to $O_3$ (Fig. 2A). Bourgeois, et al. [17] used the same observations to calculate $O_3$ contribution, of which the algorithm is described in detail in *SI Appendix*, Sect. S5 and referred as the OT method in the main text. The maximum predicted $O_{3,BB}$ and $O_{3,FF}$ values (29.8 ppb and 39.2 ppb, respectively) were substantially lower than the >100 ppb values in the training dataset, indicating that the predictions are unlikely to involve extrapolation.

**Data and materials availability**

ATom observations are available at https://doi.org/10.3334/ORNLDAAC/1581. Source code of GEOS-Chem 13.3.4 is available at https://doi.org/10.5281/zenodo.5764874.

**Acknowledgments** This study is supported by the Max Planck Society (MPG).

**Author Contributions** Y.C. conceptualized the study, procured funding and administrate the project. Y.C. and H.S. led the study. C.M., Y.C., and H.S. carried the research and wrote the paper. C.M. analyzed and visualized the data.

**Competing Interest Statement** Authors declare that they have no competing interests

***Correspondence and requests for materials should be addressed to:** H.S. (suhang@mail.iap.ac.cn) or Y.C. (yafang.cheng@mpic.de)

**Supplementary Information is available for this paper.**

# Supporting Information for

Deep learning reveals a stronger fossil fuel influence than biomass burning in shaping remote tropospheric ozone

Chaoqun Ma, Hang Su, Yafang Cheng

Corresponding author: H.S. (suhang@mail.iap.ac.cn) or Y.C. (yafang.cheng@mpic.de)

**This PDF file includes:**

Supporting text
Figures S1 to S18
SI References

## Supporting Information Text

### S1. Previous studies about tropospheric $O_3$ attribution

Source attribution of tropospheric $O_3$ typically relies on two commons approaches: perturbation or tagging.

With the perturbation approach, the contribution of a specific emission sector is defined as the change in $O_3$ mixing ratio when that sector's emissions are removed from the model. For example, to quantify the $O_3$ contribution by FF sources, two simulations are compared: one with all emissions included and one with FF emissions excluded. The difference in simulated $O_3$ between these two scenarios represent the contribution from FF sources. Because of its direct interpretability, this approach is widely used to assess the effectiveness of emission-control strategies.

The tagging approach tracks the chemical formation pathways of $O_3$ and assigns source labels to newly formed $O_3$ molecules accordingly to predefined criteria. At a given time and location, the fraction of newly produced $O_3$ attributed to BB can be defined in proportion to BB contribution to local $NO_x$ [1],VOCs [2], or even to the total air mass [3]. Approaches that estimate global annual $O_3$ production budgets [4,5] can also be broadly viewed as belonging to the tagging framework.

If $O_3$ were produced linearly from its precursors, analogous to sulfate formation from sulfur dioxide, these two approaches would yield nearly identical results. However, tropospheric $O_3$ formation is governed by highly nonlinear photochemical reactions involving $NO_x$ and VOCs. As a result, the tagging and perturbation approaches, although both conceptually valid, can produce substantially different attribution outcomes, with tagging typically yielding higher source contributions than perturbation [1]. It is therefore essential to distinguish between these two approaches when interpreting $O_3$ attribution results.

**Tagging approach**. For FF contribution to global tropospheric $O_3$ column, Grewe et al. [1] estimated a value ~30% based on their CTM calculations. Naik, et al. [4] used model to calculate the change of annual $O_3$ production from preindustrial (PI) to present day (PD) and attributes 60% (444 to 734 Tg $yr^{-1}$) net production and 42% (2391 to 5753 Tg $yr^{-1}$) chemical production of tropospheric $O_3$ to industrialization, mostly assumed to be FF contribution. With observed tracers, however, the estimation of the original tracer method ("OT method" hereafter) from Bourgeois, et al. [3] amounts to around 9% contribution for global remote troposphere according to our reproduction.

For BB contribution to global tropospheric $O_3$ column, Grewe et al. [1] yielded ~15% with the same model approach. Mao, et al. [6] found an increase of 328 Tg $yr^{-1}$ in simulated chemical production of tropospheric $O_3$ after BB emission was scaled from 0 to 1, which amounts to a contribution of nearly 6%. With the same observation dataset, we can also retrieve the contribution to remote troposphere from OT method that amounts to ~23%. Jaffe and Wigder [5] estimated about 174 Tg $yr^{-1}$ production of global tropospheric $O_3$ from BB according to the global CO emission and the $\Delta O_3/\Delta CO$ value observed from BB plumes; Jaffe and Wigder [5] considered 174 Tg $yr^{-1}$ as chemical production which amounts to a contribution of 3.5%, but the contribution could be 29% if 174 Tg $yr^{-1}$ is net chemical production of $O_3$.

**Perturbation approach**. Using a perturbation approach in their CTM simulations, Grewe et al. [1] estimated the FF contribution to global tropospheric $O_3$ column to be approximately 10%. Revell, et al. [7] simulates the change of $O_3$ from 1960 to PD and results indicate the FF contribution to global tropospheric $O_3$ at 500 hpa level could be larger than 15%. Naik, et al. [4] simulates the change of $O_3$ from PI to PD and the results show about 29% (103 Tg to 360 Tg) contribution to global tropospheric $O_3$ burden. It is hard to do emission perturbation in real word, but we could deduce the results from long-term $O_3$ observation from PI to PD at background stations. A review of $O_3$ records from remote stations shows $O_3$ level almost doubled from PI to PD [8], which means ~25% $O_3$ today around those stations is contributed by FF if we assume half of $O_3$ increase is caused by climate change.

For BB contribution to global tropospheric $O_3$ burden, we could only resort to CTM simulation. The contribution percent is about 3% (only Northern Hemisphere), 5% and 5.3% calculated from results of Pfister, et al. [9], Grewe et al. [1] and Mao, et al. [6], respectively.

Fig. S10 summarizes the spread of $O_3$ attribution results discussed above. It shows that CTM simulations generally yield a higher contribution from FF sources than from BB, whereas observation-based methods do not provide a coherent consensus on their relative importance.

## S2. Potential limitations of the OT method

After revisiting the OT method applied in Bourgeois et al. (3), we identified four main aspects that may introduce bias and potentially undermine its accuracy in assessing the $O_3$ source contribution. Besides the strong sensitivity to the differences in tracer lifetimes discussed in the main text (further extended the discussion here in the following subsection A), the remaining three issues concern the interpretation of fractional influence (subsection B), the determination of background levels (subsection C), and the weighting applied when averaging the results (subsection D). The DL framework overcomes these issues by leveraging ensemble CTM simulations to capture complex chemistry, atmospheric transport, and varying emission conditions, thereby avoiding assumptions about tracer lifetimes and background levels.

**Strong sensitivity to the differences in tracer lifetime.** According to equations [1], [2] and [3] in Bourgeois, et al. [3], the $O_3$ attributed to BB source of an air mass ($O_{3,\mathrm{BB}}$) is expressed as

$$O_{3,\mathrm{BB}} = \Delta O_3 F_{\mathrm{BB}} = \Delta O_3 \frac{\Delta HCN}{\Delta CO \cdot ER_{\mathrm{BB}}}, \qquad \text{[Eq. S1]}$$

where $\Delta HCN$, $\Delta CO$ and $\Delta O_3$ are the excess mixing ratios of HCN (BB-specific tracer), CO and $O_3$, which are the difference between the mixing ratio of these compounds and their background levels. The background level, by its definition, should represent the mixing ratio without any influence of BB or FF sources. $ER_{\mathrm{BB}}$ is the emission ratio of HCN to CO in the BB emission inventory. $F_{\mathrm{BB}} = \frac{\Delta HCN}{\Delta CO \cdot ER_{\mathrm{BB}}}$ is then the fractional influence of BB to that air mass. The background level here should represent the level when the airmass is not influenced by both BB and FF emission.

$O_{3,\mathrm{FF}}$ can be derived analogously, except that HCN is replaced by an FF-specific tracer. In our case, $C_2Cl_4$ is the FF-specific tracer and $O_{3,\mathrm{FF}} = \Delta O_3 F_{FF} = \Delta O_3 \frac{\Delta C_2Cl_4}{\Delta CO \cdot ER_{\mathrm{FF}}}$, where $ER_{\mathrm{FF}}$ is the emission ratio of $C_2Cl_4$ to CO in the FF emission inventory.

However, as discussed in the main text, $F_{\mathrm{BB}}$ could be interpreted as the fraction of BB influence only if CO and HCN decay at the same rate after emitted from the source. With a shorter lifetime of CO than HCN in reality, an air parcel will see $F_{BB}$ constantly increase after it leaves the source region even no additional $O_3$ production or depletion. The effect is particularly pronounced in regions far from the source after long transit time, e. g., over Southern Hemisphere extratropical Atlantic.

Using the link between the $F_{BB}$ and the transit time, we can estimate the transit time *t* of an air mass. Following the method (Eq. S5 and S6) in Bourgeois, et al. [3], the BB-induced enhancements of HCN and CO can be approximated as:

$$\Delta HCN_{\mathrm{BB}} = c_{\mathrm{BB,HCN}} \exp\left(\frac{-t}{\tau_{\mathrm{HCN}}}\right), \quad \text{[Eq. S2]}$$

$$\Delta CO_{\mathrm{BB}} = c_{\mathrm{BB,CO}} \exp\left(\frac{-t}{\tau_{\mathrm{CO}}}\right) \quad \text{[Eq. S3]}$$

Here, $t$ is the transit time of the air mass; $c_{\mathrm{BB,HCN}}$ ($c_{\mathrm{BB,CO}}$) are the mixing ratio of HCN (CO) after freshly emitted by BB; $\tau_{\mathrm{HCN}}$ and $\tau_{\mathrm{CO}}$ are the lifetimes for HCN and CO. Dividing Eq. S3 by Eq. S2 yields:

$$\frac{\Delta HCN_{\mathrm{BB}}}{\Delta CO_{\mathrm{BB}}} = ER_{\mathrm{BB}} \exp\left[\left(\frac{1}{\tau_{\mathrm{CO}}} - \frac{1}{\tau_{\mathrm{HCN}}}\right) t\right]. \quad \text{[Eq. S4]}$$

Because $\Delta HCN$ can only be contributed by BB sources, then $\Delta HCN_{\mathrm{BB}} = \Delta HCN$. By defining $R_{\mathrm{BB}}$ as the fractional contribution of BB to $\Delta CO$, we have $\Delta CO_{\mathrm{BB}} = \Delta CO \cdot R_{\mathrm{BB}}$. Recalling the definition $F_{\mathrm{BB}} = \frac{\Delta HCN}{\Delta CO \cdot ER_{\mathrm{BB}}}$, the transit time $t$ can be solved from Eq. S4 as

$$t = ln\left(\frac{F_{\mathrm{BB}}}{R_{\mathrm{BB}}}\right) / (1/\tau_{\mathrm{CO}} - 1/\tau_{\mathrm{HCN}}), \quad \text{[Eq. S5]}$$

Here, we adopted 150 days for $\tau_{\mathrm{HCN}}$ and 60 days for $\tau_{\mathrm{CO}}$ [10,11] and tested the full range of previously reported $R_{\mathrm{BB}}$ from 33% to 91% [3] over Southern Hemisphere extratropical Atlantic during austral summer & spring. We then estimate a transit time of about 83-184 days in this region and season.

The estimated transit time of 83-184 days is substantially longer than the 2-28 days inferred from backward trajectory analyses used by Bourgeois, et al. [3]. Trajectory-based analyses may underestimate the true transit time of an air parcel when it has experienced atmospheric mixing with more aged air along its path or undergone earlier transport that trajectories fail to capture. The mixing processes are driven by turbulence and entrainment/detrainment in convection and their effects cannot be well represented in standard trajectory simulations [12,13]. In addition, trajectory-based age is defined as the time required for the air parcel to travel backward from the target location to its first interception with the land boundary layer or with rising fire plumes [14]. However, air within the boundary layer or fire plumes may already undergo mixture with non-freshly-emitted air.

As an attempt to illustrate the impact of this bias (Fig.1B), we tried to correct the $F_{\mathrm{BB}}$ according to the CO-HCN lifetime difference by adjusting the HCN mixing ratio. Assuming an exponential decay function, the corrected $F_{BB}^{*}$ is

$$F_{\mathrm{BB}}^{*} = \left(\frac{\Delta HCN}{c_0}\right)^{\frac{\tau_{\mathrm{HCN}}}{\tau_{\mathrm{CO}}}} / \frac{\Delta CO \cdot ER_{\mathrm{BB}}}{c_0}, \quad \text{[Eq. S6]}$$

where $c_0$ is the mixing ratio of $HCN$ after freshly emitted by BB.

The value of $c_0$ was approximated here as the near-surface $\Delta HCN$ averaged over the source region. For the tropical and southern extratropical Atlantic, potential HCN sources include biomass burning hotspots in South America, Africa, and Australia, with corresponding $c_0$ values of 383, 914, and 314 ppt, respectively (Fig. S11). For simplicity, we used the average of the three initial values (383, 914, and 314 ppt) and assign $c_0$ as 503 ppt, which results in the distribution of $F_{\mathrm{BB}}^{*}$ as shown in Fig. S2D and the corresponding $O_{3,BB}$ in Fig. 1C. The revised results show higher $O_{3,BB}$ near the most intensified BB source regions over tropical Atlantic, indicating the lifetime difference contributes to the biases observed in the OT methods. The current correction in Fig. 1C is intentionally simplified to provide an intuitive demonstration of the effect of tracer lifetime differences, particularly in the estimation of $c_0$. A more refined and physically and chemically grounded treatment of tracer lifetime effects is achieved in this study through a deep learning–based framework that integrates observational constraints with source–tracer relationships learned from chemical transport modeling. This analysis highlights the importance of explicitly accounting for tracer lifetime differences when applying observation-based tracer approaches to source attribution in the remote troposphere.

**Interpretation of fractional influence.** Following Eq. S12 in Bourgeois, et al. [3], the fractional influence of FF source to an air mass is $F_{\mathrm{FF}}$ and can be expressed as

$$F_{\mathrm{FF}} = \frac{\Delta FF}{\Delta CO \cdot ER_{\mathrm{FF}}} = \frac{\sum_i X_i^{\mathrm{FF}} \times FF_i^{\mathrm{S}}}{ER_{\mathrm{FF}} \times \left(\sum_i X_i^{\mathrm{FF}} \times CO_i^{\mathrm{S}} + \sum_j X_j^{\mathrm{BB}} \times CO_j^{\mathrm{S}}\right)}, \qquad \text{[Eq. S7]}$$

where $ER_{\mathrm{FF}}$ is the emission ratio between the FF-specific tracer (like $C_2Cl_4$ in this work) and CO in the FF emission. $\mathrm{i}$ indicates the $\mathrm{ith}$ air parcel from FF while $\mathrm{j}$ indicates the $\mathrm{jth}$ air parcel from BB. $CO_i^{\mathrm{S}}$ and $CO_j^{\mathrm{S}}$ are the mixing ratios of CO at BB and FF sources respectively. $FF_i^{\mathrm{S}}$ is the mixing ratio of FF-specific tracer at FF source (like $C_2Cl_4$ in this work). $X_i^{\mathrm{FF}}$ is the contribution of the $\mathrm{ith}$ FF air parcel to the observed air mass and $X_j^{\mathrm{BB}}$ is the contribution of the $\mathrm{jth}$ BB air parcel to the observed air mass. Assuming a constant $ER_{FF}$ for each air parcel $\mathrm{i}$ , this expression could be arranged to

$$F_{\mathrm{FF}} = \frac{\sum_i X_i^{\mathrm{FF}} \times CO_i^{\mathrm{S}}}{\sum_i X_i^{\mathrm{FF}} \times CO_i^{\mathrm{S}} + \sum_j X_j^{\mathrm{BB}} \times CO_j^{\mathrm{S}}}. \qquad \text{[Eq. S8]}$$

This quantity represents the fraction of CO originating from FF emissions within the portion of CO that exceeds the background level, yet in the OT method it is assumed to equal the fractional contribution to $O_3$. However, the relationship between this CO-based fraction and $O_3$ contribution depends on complex atmospheric processes like emission, transportation and nonlinear $O_3$ chemistry. Even ignoring the nonlinearity of $O_3$ chemistry and the potential differences in environmental conditions during transport (e.g., radiation and temperature), the two would only be equivalent if FF and BB have the same profile of emission factors. In reality, this assumption does not hold. (1) Air parcel emitted by FF source usually has lower mixing ratio of CO than that from BB because anthropogenic combustion of fuels [15] usually has higher combustion efficiency than BB processes [16]; (2) Air parcels from BB and FF sources also differ in their $O_3$ production potential due to variations in chemical composition—for example, different $NO_x$ concentrations.

To correct this bias, we assume the potential of $O_3$ production for each air parcel is proportional to its $NO_x$ concentration when emitted. That is based on the assumption of $NO_x$-limited $O_3$ production for remote tropospheric region [6,17]. Then, the "corrected" fraction of $O_3$ contribution could be written as $F_{\mathrm{FF}}^*$ and

$$F_{\mathrm{FF}}^* = \frac{\sum_i X_i^{\mathrm{FF}} \times CO_i^{\mathrm{S}} \times \frac{NO_{x\mathrm{FF}}^{\mathrm{S}}}{CO_{\mathrm{FF}}^{\mathrm{S}}}}{\sum_i X_i^{\mathrm{FF}} \times CO_i^{\mathrm{S}} \times \frac{NO_{x\mathrm{FF}}^{\mathrm{S}}}{CO_{\mathrm{FF}}^{\mathrm{S}}} + \sum_j X_j^{\mathrm{BB}} \times CO_j^{S} \times \frac{NO_{x\mathrm{BB}}^{\mathrm{S}}}{CO_{\mathrm{BB}}^{\mathrm{S}}}}. \qquad \text{[Eq. S9]}$$

That could be further reduced to the following if we assume $r = \frac{NO_{x\mathrm{FF}}^{\mathrm{S}}}{CO_{\mathrm{FF}}^{\mathrm{S}}} \Big/ \frac{NO_{x\mathrm{BB}}^{\mathrm{S}}}{CO_{\mathrm{BB}}^{\mathrm{S}}} = \frac{CO_{\mathrm{BB}}^{\mathrm{S}}}{NO_{x\mathrm{BB}}^{\mathrm{S}}} \Big/ \frac{CO_{\mathrm{FF}}^{\mathrm{S}}}{NO_{x\mathrm{FF}}^{\mathrm{S}}}$

$$F_{\mathrm{FF}}^* = \frac{\sum_i X_i^{\mathrm{FF}} \times CO_i^{\mathrm{S}}}{r \times (\sum_i X_i^{\mathrm{FF}} \times CO_i^{\mathrm{S}}) + \sum_j X_j^{\mathrm{BB}} \times CO_j^{\mathrm{S}}}. \qquad \text{[Eq. S10]}$$

With some simplification, we can get

$$F_{\mathrm{FF}}^* = \frac{1}{1 - \frac{1}{r} + \frac{1}{rF_{\mathrm{FF}}}}. \qquad \text{[Eq. S11]}$$

With the same calculation, similar correction could be obtained for $F_{BB}$ as

$F^{*}_{\mathrm{BB}} = \frac{1}{1-r+\frac{r}{F_{\mathrm{BB}}}}$. [Eq. S12]

The value of $c$ could range from 4.4 ($\frac{CO^{\mathrm{S}}_{\mathrm{BB}}}{NO^{\mathrm{S}}_{x\mathrm{BB}}}$ is 29.6 ppb ppb$^{-1}$ for grass and savanna [16] and $\frac{CO^{\mathrm{S}}_{\mathrm{FF}}}{NO^{\mathrm{S}}_{x\mathrm{FF}}}$ is 6.75 ppb ppb$^{-1}$ by inspecting the total anthropogenic emission of China 2016 (http://www.meicmodel.org, MEIC)) to 16.3 ($\frac{CO^{\mathrm{S}}_{\mathrm{BB}}}{NO^{\mathrm{S}}_{x\mathrm{BB}}}$ is 110 ppb ppb$^{-1}$ for boreal forest [16] and $\frac{CO^{\mathrm{S}}_{\mathrm{FF}}}{NO^{\mathrm{S}}_{x\mathrm{FF}}}$ is still 6.75 ppb ppb$^{-1}$). Because $c$ is usually larger than 1, the corrected fraction $F^{*}_{\mathrm{BB}}$ ($F^{*}_{\mathrm{FF}}$) is always smaller (larger) than the original $F_{\mathrm{BB}}$ ($F_{\mathrm{FF}}$). This indicates that the original evaluation systematically overestimated (underestimated) the contribution from BB (FF).

To check how large the change could be after correction, we adopted the value 16.3 for $c$ and replaced $F_{\mathrm{BB}}$ ($F_{\mathrm{FF}}$) by $F^{*}_{\mathrm{BB}}$ ($F^{*}_{\mathrm{FF}}$) during subsequent reproduction of the OT method. The resulting distribution (Fig. S12B) shows that, although not identical, the corrected data points fall within a region closer to that predicted by our DL model (Fig. 4A) and indicates a generally higher contribution from FF emissions compared with BB sources.

To further examine the impact of different emission ratios of $NO_x$ to CO, we conducted an additional CTM simulation scenario ("SameER"), in which BB and FF sources were assumed to share identical emission ratio profiles. This scenario is identical to the Base simulation, except that $NO_x$ emissions from both BB and FF sectors were prescribed as a fixed ratio to CO emissions, rather than following the original $NO_x$ emission inventories. The emission ratio is chosen to be 0.148 ppb NOx per ppb CO (a typical value for anthropogenic emissions) for both FF and BB sectors. Then, a DL model was trained only with simulations from the SameER to predict the contribution. To estimate the uncertainty associated with the choice of emission ratio, we also tested emission ratios varied by a factor of two (Fig. S13B).

As shown in Fig. S13A, all data points (triangles) shift toward higher BB contributions compared with the default results (filled circles). Regarding the ratio of $O_{3,\mathrm{BB}}$ to $O_{3,\mathrm{FF}}$, DL model trained on SameER data generally yields results that agree better with Bourgeois, et al. [3] than the default DL model trained on simulations with realistic, source-dependent emission ratios. In Trop ATL, NHE S&S, SHE S&S and SHE F&W regions, where the default DL model used to predict higher $O_3$ from FF than BB, the SameER-trained DL model now predicts the opposite pattern.

**Approximation of background levels.** In the OT method, the background levels of the tracers CO (HCN, $C_2Cl_4$) are required to calculate the excess mixing ratios $\Delta CO$ ($\Delta HCN$, $\Delta C_2Cl_4$) and then the influence fraction $F_{\mathrm{BB}}$ and $F_{\mathrm{FF}}$. The background level, by its definition, should represent the mixing ratio without any influence of BB or FF sources.

However, in the OT method, the background is approximated as the average of air masses with mixing ratios of both HCN and $C_2Cl_4$ below the regional median values. This approach does not guarantee that $\Delta CO$ remains positive, resulting in numerous negative or near-zero $\Delta CO$ values. Since $\Delta CO$ appears in the denominator of Eqs. S1 and S3, such cases produce unrealistically high or negative $F_{\mathrm{BB}}$ and $F_{\mathrm{FF}}$ values. These outliers must then be excluded before averaging, in order to obtain results comparable to those shown in their Fig. 4. This may partly explain their weak correlation between the resultant $O_3$ attribution and its tracer, which should theoretically be much stronger as demonstrated by our DL method (Fig. S14).

In this study, when replicating the OT method using simulated inputs (Figs. 1, 2, S1, and S2), the background levels are instead determined from simulations with both BB and anthropogenic emissions removed to achieve a smooth distribution of results.

**Weighting applied during averaging the results.** After obtaining the $O_3$ attribution for each measurement (eq. (3) of Bourgeois, et al. [3]), they averaged the results with the mixing ratio of the corresponding tracer as a weighting factor (eq. (4) of Bourgeois, et al. [3]). This weighting is unnecessary because, if each measurement is assumed to represent an equal spatial extent, all observations should contribute equally to the average. Applying this weight artificially amplifies the difference between BB and FF contribution, because a measurement with higher BB (FF) contribution also likely to exhibit higher HCN ($C_2Cl_4$) mixing ratio and thus larger weights.

After removing the weighting during averaging (Fig. S12A), the mean results generally get closer to the 1:1 line. However, the results still indicate equal or larger contribution from BB than FF sources, suggesting that this weighting issue is only a minor factor contributing to the discrepancies in $O_3$ attribution.

### S3. Impact of uncertainties in $O_3$ chemistry to $O_3$ contribution from biomass burning

Here we carried a new CTM simulation scenario (maxBBp) to maximized the $O_3$ production from unit biomass burned according to known uncertainties in model $O_3$ chemistry. The configuration is the same as Base except for the following three aspects:

(i) The uptake coefficient $\gamma$ of hydroperoxyl ($HO_2$) on aerosols. The heterogeneous uptake of $HO_2$ on aerosol may suppress ozone formation. $\gamma$ is set to 0.2 in default but has been reported to vary from 0 to 0.4 [18,19]. To represent the lower bound ($\gamma \approx 0$), primary aerosol emissions from BB sources were turned off, thereby enhancing $O_3$ production from BB.

(ii) Nitrous acid (HONO), terpene and isoprene emissions. They are not included in the QFED2 inventory but could contribute to the $O_3$ formation. We thus added them according to the highest emission factors reported [16,20].

(iii) Furan and some other VOCs emitted by BB, which are not included in the GEOS-Chem chemistry scheme, might contribute significantly to OH reactivity. To account for this, we scaled the VOC emission from QFED2 by a factor of 2 according to the reported potential underestimation of OH reactivity due to these missing species [21]

The 3 measures above were deliberately designed to favor $O_3$ contribution from BB because we intend to show $O_3$ contribution from BB is still less than that from FF after accounting for this extreme case. For the same reason, we did not compensate the possible overestimation of BB $NO_x$ [22] because that would make BB produce less $O_3$.

Fig. S4 shows the simulated contribution of BB relative to FF sources from the maxBBp scenario. Compared with the Base scenario, the relative importance of BB increases but remains well below that of FF sources. With ATom observations as input, our DL model predicts results similar to the Base scenario. Therefore, the uncertainties of $O_3$ chemistry in BB smoke is unable to change the conclusion that FF source contributes more to $O_3$ in the remote troposphere than BB.

### S4. Sensitivity of the $O_3$ contribution to emission ratios of tracers

This section will test how sensitive our DL framework is to the uncertainties in tracer emission ratio. Considering HCN ($C_2Cl_4$), as a tracer of BB (FF) emissions, plays an important role in DL model prediction (*SI Appendix,* Sect. S6), change of the emission ratios for HCN and $C_2Cl_4$ (5.7 pptv ppbv$^{-1}$ of BB CO and 0.03 pptv ppbv$^{-1}$ of anthropogenic CO emission for default) is sure to change the attribution. Generally, overestimation of HCN ($C_2Cl_4$) emission ratios in the model will make the CTM produce less $O_{3,BB}$ ($O_{3,FF}$) per unit of HCN ($C_2Cl_4$); such error, after learned by the DL model, will cause underestimation of the $O_{3,BB}$ ($O_{3,FF}$) given the same observed HCN ($C_2Cl_4$).

To evaluate the sensitivity of the average $O_3$ contribution to the change of emission ratios, we consider a case extremely favors BB ozone contribution in which we assume the "correct" emission ratio for HCN ($C_2Cl_4$) is 0.8 (1.2) times of our default configuration. Because both HCN and $C_2Cl_4$ are set as a passive tracer which does not influence other processes, previous simulation results could be "corrected" by simply scaling the mixing ratio of HCN and $C_2Cl_4$ by 0.8 and 1.2 respectively without rerunning the CTM. The subsequent training and analysis procedures, consistent with those in Material and Methods, were then repeated.

As expected, the "corrected" DL model predicts higher (lower) $O_3$ contributed by BB (FF) emissions than the default (Figs. S15A and S15B) and the averaged results also assign more importance to the $O_{3,BB}$ (Fig. S15C). However, the relative change (~8%) is much smaller than the perturbation in tracer emission ratios (~20%). This indicates that The DL model is not overly sensitive to uncertainties in these tracers. Its robustness is reinforced by the fact that the DL framework implicitly captures influence of the nonlinear $O_3$-precursors relationships governed by chemistry and transport, processes for which GEOS-Chem has long demonstrated strong performance [23,24].

In addition, the "corrected" results likely exceed the realistic upper limit of $O_{3,BB}$, as our Base simulation already slightly underestimates observed HCN (and overestimates $C_2Cl_4$) with minimal bias in CO (Fig. S5). Therefore, further reducing HCN (increasing $C_2Cl_4$) by applying a 0.8 (1.2) correction factor likely would likely drive the simulation further from, rather than closer to, reality.

### S5. Impact of non-linear chemistry to CTM simulated $O_3$ contribution

To train the DL model, we estimated the $O_3$ contribution from a specific emission source as the difference between CTM simulations with all emissions and those with that emissions source excluded. This "subtracting approach" is commonly used to represent the marginal effect of individual emission sectors but can be biased due to the nonlinear chemistry of $O_3$. As discussed in Grewe et al. [1], the same increase in $NO_x$ emissions can produce a substantial $O_3$ increase under low-$NO_x$ conditions but little or no increase under high-$NO_x$ conditions—a saturation effect already observed at many Northern Hemisphere continental sites. Natural background $NO_x$ sources, including lightning and soil $NO_x$, are included in GEOS-Chem via parameterizations detailed in Murray *et al.* [25] and Weng, et al. [26].

To assess the impact of this nonlinearity, we conducted an additional set of simulations similar to the Base scenario: (i) excluding both BB and anthropogenic emissions, (ii) excluding only BB, and (iii) excluding only anthropogenic emissions. In this "adding approach," $O_3$ contributions from BB and FF sources are computed as the differences between (iii) and (i), and (ii) and (i), respectively—opposite in direction to the subtracting approach.

As shown in Figs. S16A and S16B, the averaged $O_{3,BB}$ and $O_{3,FF}$ determined by adding approach has almost the same spatial distribution as that by subtracting approach (Fig. S6). The adding approach generally yields slightly higher contributions, but the differences are modest and similar for $O_{3,BB}$ and $O_{3,FF}$ (4% higher for $O_{3,BB}$, 5% higher for $O_{3,FF}$; Fig. S16C and S16D). These results suggest that nonlinear effects have minimal impacts to the $O_3$ contribution simulated by CTM in the main text. In addition, this uncertainty is insufficient to alter our main conclusion that FF contribute more tropospheric $O_3$ in remote area than BB.

## S6. Interpretation of deep learning model

In this section, we tried two interpretation techniques (or "XAI" in another word) to probe how the trained DL model is working and whether it makes prediction according to reasonable relationships.

**SHAP value.** The SHAP (SHapley Additive exPlanations) value is a method based on cooperative game theory that provides interpretable explanations of model predictions. It assigns an importance value to each feature (i.e., input variable in this study), representing the marginal effect of including that feature on the model's output [27,28].

As an additive feature attribution approach, SHAP provides a unique decomposition of the model prediction for a specific input sample. This decomposition attributes the deviation of the model output from its expected value ($\mathrm{M}(x_0) - \mathrm{E}[\mathrm{M}(X)]$), to each input feature. M() denotes the model, $x_0$ is the current input, $X$ is the set of all inputs, and E[] is the expectation. Features with positive SHAP values contribute to higher-than-average model predictions, while negative SHAP values indicate a suppressing effect. Features with SHAP values near zero contribute little to the prediction.

The SHAP framework satisfies three desirable properties—local accuracy, missingness, and consistency. Among these, local accuracy ensures that the sum of all feature attributions equals the model output for the given instance, regardless of model linearity.

To calculate the SHAP value, we need a set of input samples $\mathrm{X}$ with a specific input $x_0$. Because the calculation time increases dramatically with the increase of the number of samples in $\mathrm{X}$, we only randomly selected 1000 samples from the $8\times10^5$ training samples and calculated the SHAP value for the input $x_0$ which generates the maximum (orange bars in Fig. S17) and minimum (blue bars in Fig. S17) $O_3$ contribution. To evaluate the uncertainty of subsampling, we repeated the process 20 times to see the standard deviation (grey bars in Fig. S17) of the results.

As shown in Fig. S17, the level of $O_{3,BB}$ and $O_{3,FF}$ both depend on the level of total $O_3$, as higher absolute contributions from BB or FF sources naturally occur when overall $O_3$ levels are elevated. Also, it is quite reasonable that $O_{3,BB}$ ($O_{3,FF}$) is related to its corresponding tracer HCN ($C_2Cl_4$) while unrelated to the other $C_2Cl_4$ (HCN). The CO mixing ratio provides more predictive power for $O_{3,BB}$ than for $O_{3,FF}$, since CO variability primarily reflects BB activity, whereas anthropogenic CO forms a relatively stable background [29,30]. The DL model places greater emphasis on geographical variables like latitude in predicting $O_{3,FF}$ because $O_{3,FF}$ is always higher in NH than in SH (Figs.3 and S6).

**Partial-dependence plots.** A commonly used and straightforward method to determine the importance of input variables is the use of partial-dependence plots. This method excludes the input variable (or predictor variable) of interest and represent the marginal effect of the variable on the prediction with the resulting reduction in performance [31]. The way to exclude a variable could vary and here we adopted the "permutation accuracy importance" measure [32] in which we randomly permuted an input variable of interest with the remaining input variables unpermuted.

Fig. S18 shows the change of $R^2$ calculated with the whole training dataset before and after permutation. This method ranks the importance of input variables in a way similar to the SHAP values: performance in predicting $O_{3,BB}$ and $O_{3,FF}$ both have large reduction after excluding total $O_3$;

$O_{3,BB}$ ($O_{3,FF}$) rely more on HCN ($C_2Cl_4$) than $C_2Cl_4$ (HCN); the performance in predicting $O_{3,BB}$ depends more on CO than FF ozone; geographical variables provide more information in predicting $O_{3,FF}$ than in $O_{3,BB}$.

We tried other commonly used interpretation technologies like backward optimization and layer relevance propagation [33-35] but they perform bad for being designed to work best for the classification instead of the regression task in our research. Given the strong consistency between the variable importance derived from the two employed methods and their alignment with established physical understanding, we conclude that our DL model provides reliable results.

## Figures

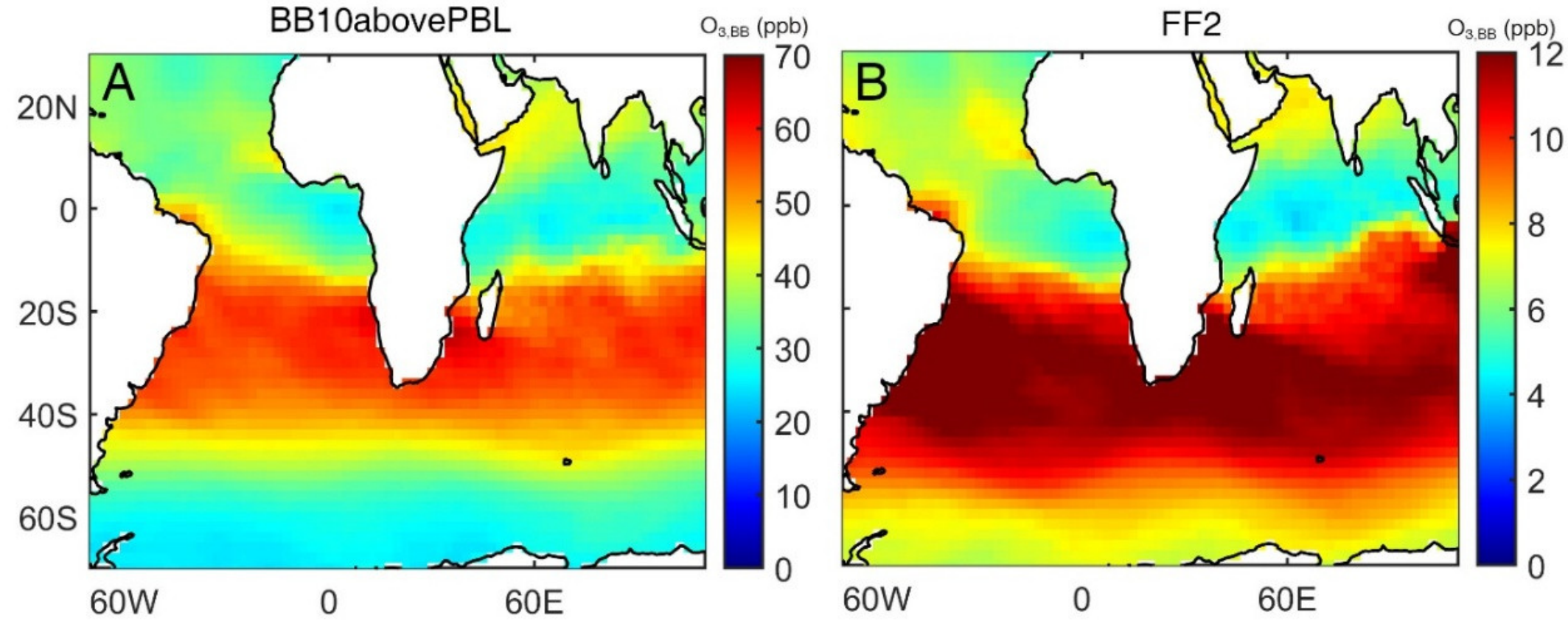

**Fig. S1. $O_3$ contributions from biomass burning (BB) estimated using the observation-based tracer approach of Bourgeois, et al. [3] (the OT method) under two alternative emission scenarios: BB10abovePBL (A) and FF2 (B).** $O_3$ contribution from BB ($O_{3,BB}$) here is at 500 hPa level over the ocean, averaged over December 2017 to January 2018. BB10abovePBL is the same as Base in Fig. 1B except BB emissions are scaled by a factor of 10 and released entirely above the planetary boundary layer (PBL). FF2 is the same as Base except anthropogenic emissions were doubled. Despite these changes, both scenarios retain the same spatial distribution of BB emissions and wind fields as in Fig. 1A, and thus $O_{3,BB}$ is expected to remain higher over the tropical Atlantic than the extratropical Atlantic, as discussed in the main text. However, the OT method again produces the same opposite pattern as the behavior seen in Fig. 1B. This recurring, counter-intuitive behavior across distinct scenarios highlights an intrinsic limitation of the OT method rather than an artifact of any specific emission configuration.

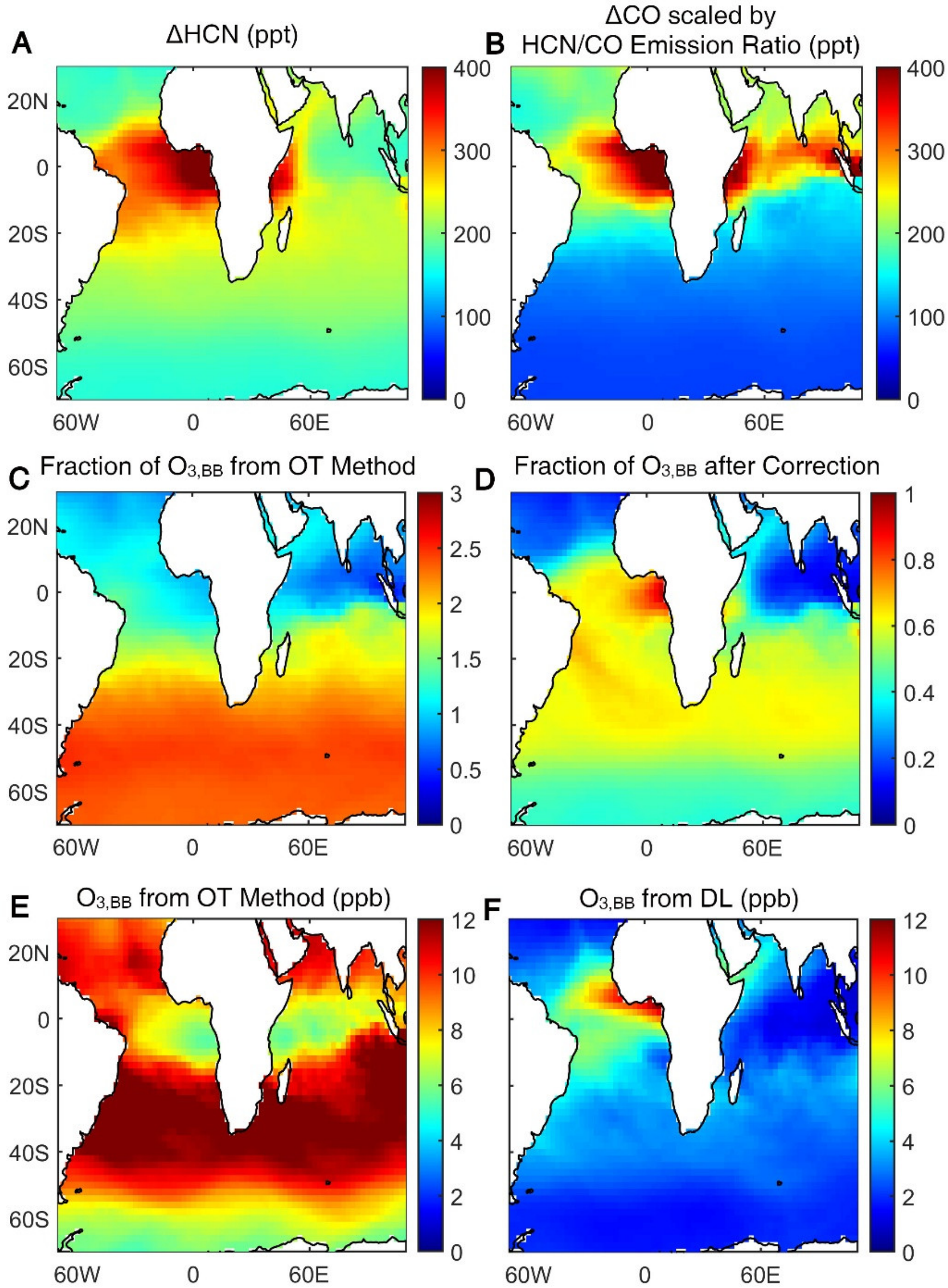

**Fig. S2. Comparison of $O_3$ contribution from biomass burning (BB) calculated by the observation-based tracer approach of Bourgeois, et al. [3] (the OT method) and our deep learning (DL) model.** (**A**) Excess hydrogen cyanide mixing ratio (ΔHCN) at 500hpa from the Base simulation scenario, averaged over December 2017–January 2018. (**B**) The same to (A), except for the excess CO mixing ratio scaled by the emission ratio of HCN to CO. (**C**) Fraction of ozone attributed to BB using the OT method, computed as the ratio of (A) to (B). (**D**) The same to (C) except the fraction was corrected according to the lifetime difference between HCN and CO (*SI Appendix,* Sect. S2). (**E**) $O_{3,BB}$ estimated by the OT method using tracers simulated by Base as input. (**F**) The same to (E) except by the DL model.

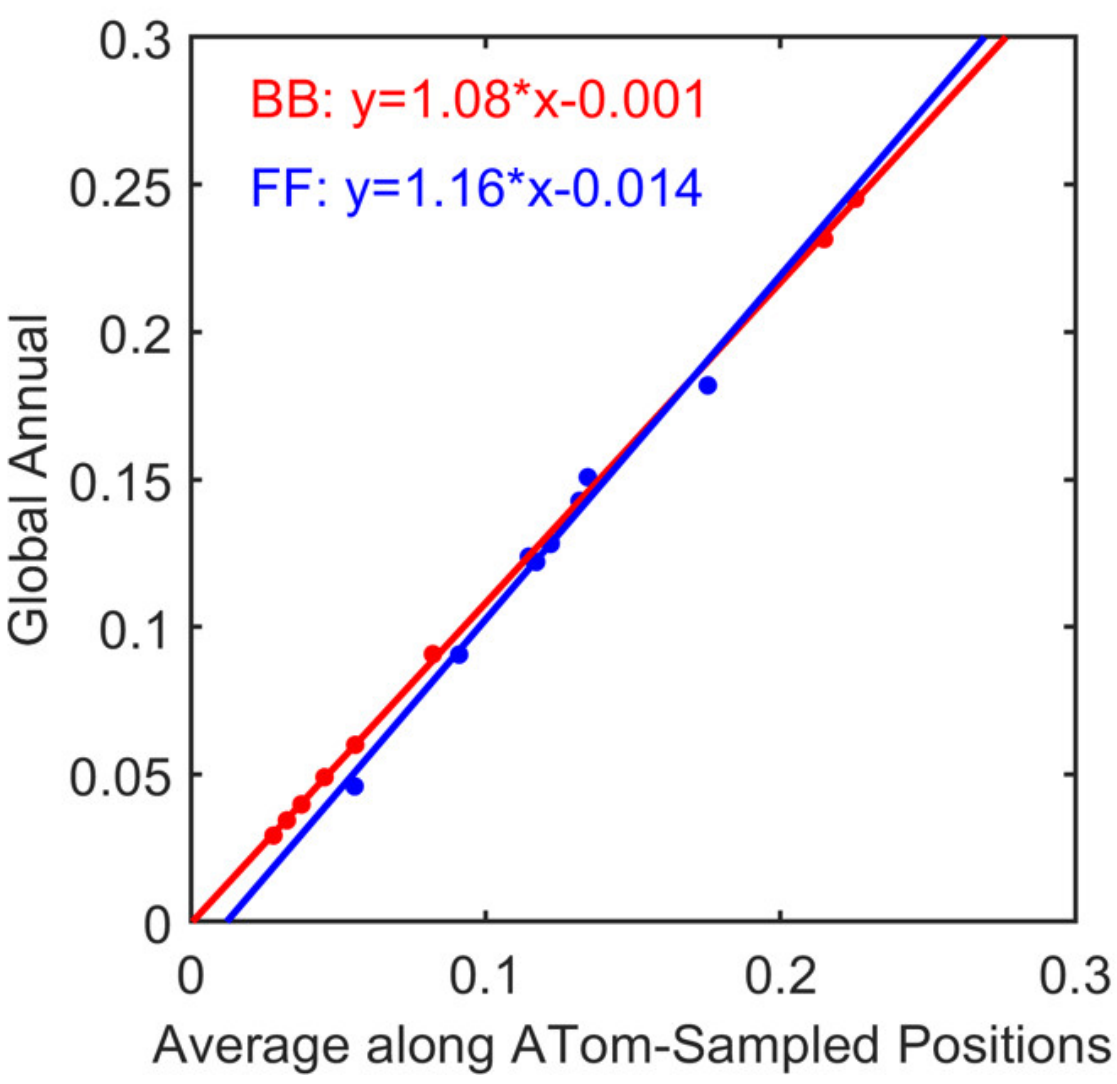

**Fig. S3. Relationship between the global annual average $O_3$ relative contribution and the average derived from ATom-sampled locations.** Each dot represents the average from an individual simulation scenario, and solid lines indicate linear regressions. For each simulation at ATom-sampled points in the remote troposphere, the $O_3$ contribution from BB or FF and the total $O_3$ mixing ratio are averaged using air density as a weight. The ATom-based relative contribution (x-axis) is computed as the ratio of the density-weighted average $O_3$ contribution to the density-weighted average total $O_3$. The global annual relative contribution (y-axis) is computed similarly but using all model grids within the remote troposphere. Differences between the x- and y-values reflect the underrepresentation of global conditions by the limited ATom sampling coverage. Global averaged $O_3$ mixing ratio from BB or FF (Fig. 4B) is calculated in a similar way except x and y axis now represent the averaged $O_3$ mixing ratio.

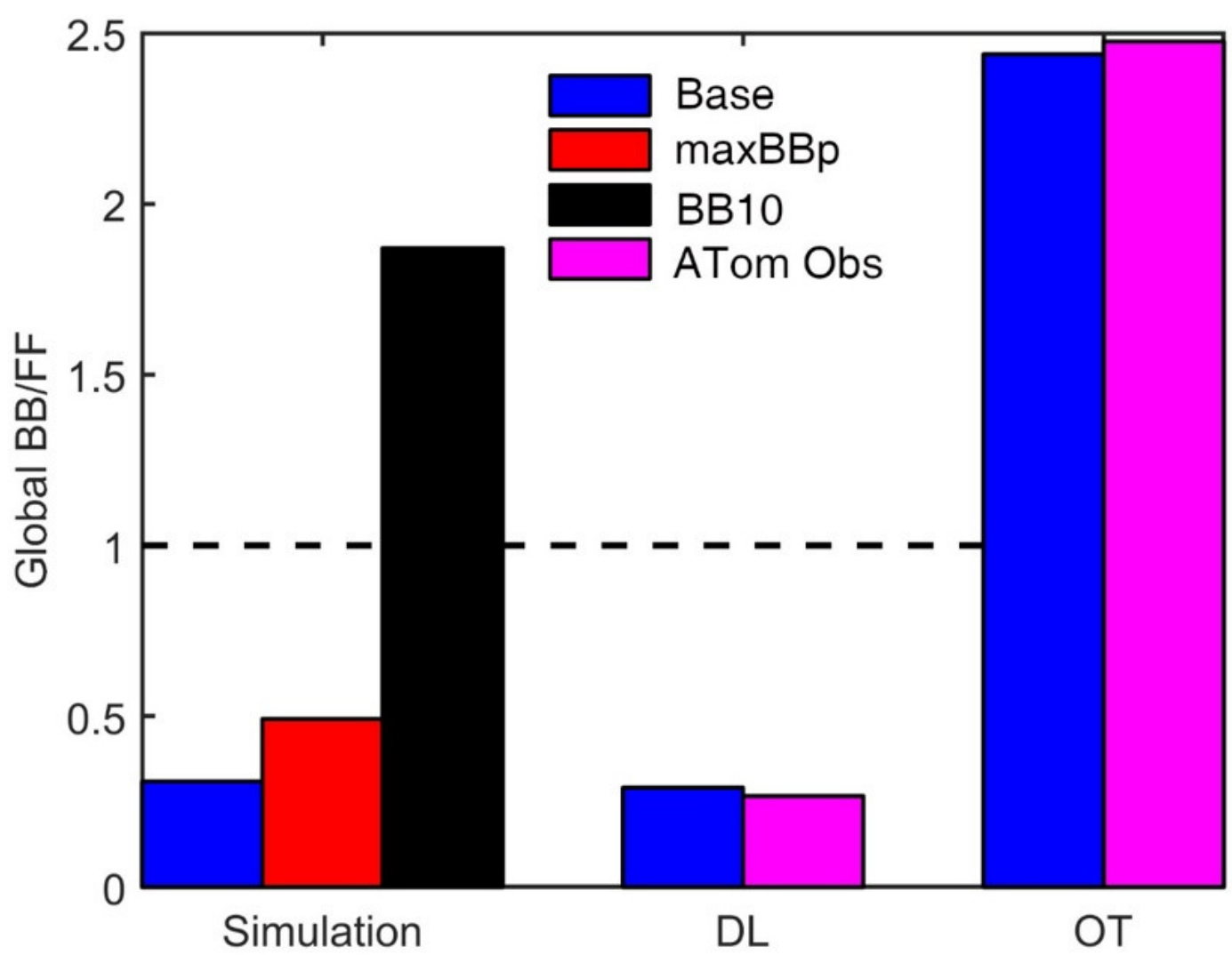

**Fig. S4. Ratio of global average $O_3$ contributions from biomass burning (BB) or fossil fuel (FF) sources.** Global average $O_3$ contributions are the relative contribution derived from the linear relationships shown in Fig. S3. Results are compared across three methods: differencing chemical transport model (CTM) simulations, the DL model, and the OT method [3], from left to right. Inputs are from Base (blue), maxBBp (red), BB10 (black) simulations and Atom observations (magenta). maxBBp is a scenario that maximizes the $O_3$ production from unit biomass burned according to known uncertainties in model $O_3$ chemistry (see Sect. S3). BB10 scenario is the same as Base except BB emission is scaled by 10, raising the BB/FF ratio from 0.31 to 1.87. This nonlinear response indicates that even a twofold increase in BB emissions is insufficient to achieve a one-to-one balance between BB and FF contributions (increase the ratio from 0.31 to 1).

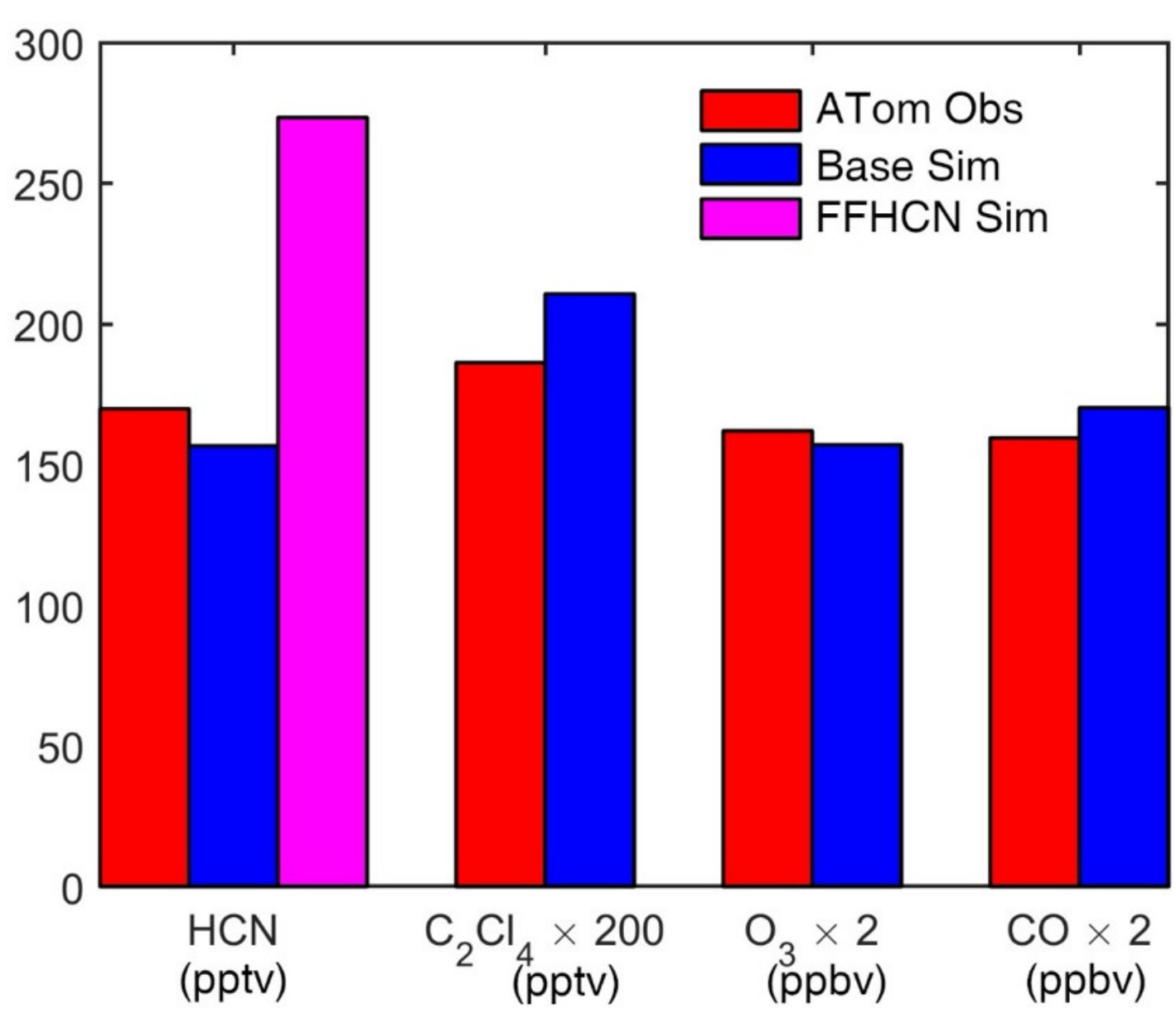

**Fig. S5. Comparison of the average mixing ratios between the ATom observations and corresponding model simulations in the remote troposphere.**

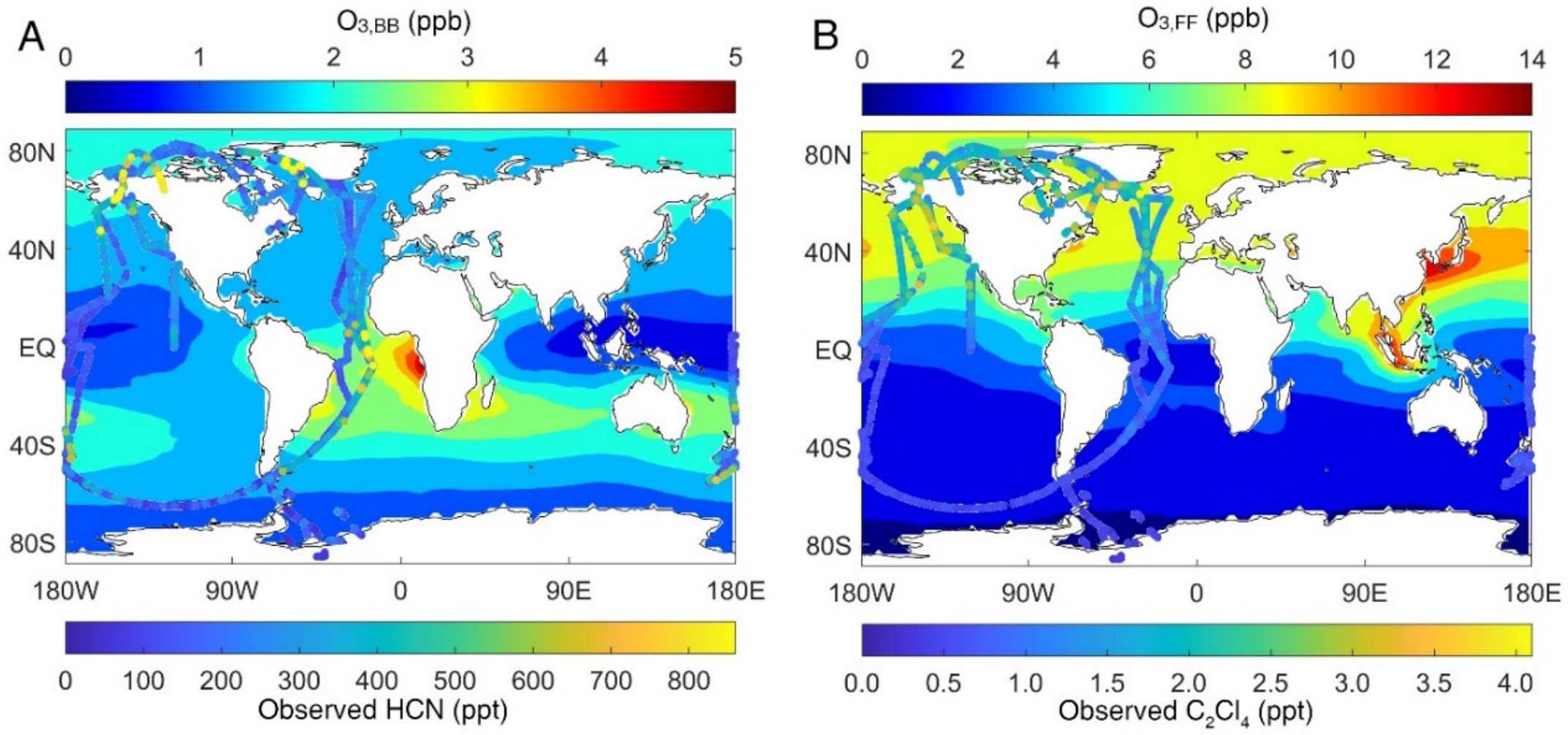

**Fig. S6. $O_3$ contribution from biomass burning (BB) and fossil fuel (FF) sources simulated by the Base scenario (color shading) and observed mixing ratio of their tracers (colored dots).** $O_3$ contributions from BB ($O_{3,BB}$) and FF ($O_{3,FF}$) are the average from June 2016 to May 2018 at the 500 hPa level. Tracers for BB and FF emissions are hydrogen cyanide (HCN) and tetrachloroethylene ($C_2Cl_4$), respectively, as measured by the ATom aircraft campaigns over four seasonal global circuits. Note the different color scales between the left and right panels.

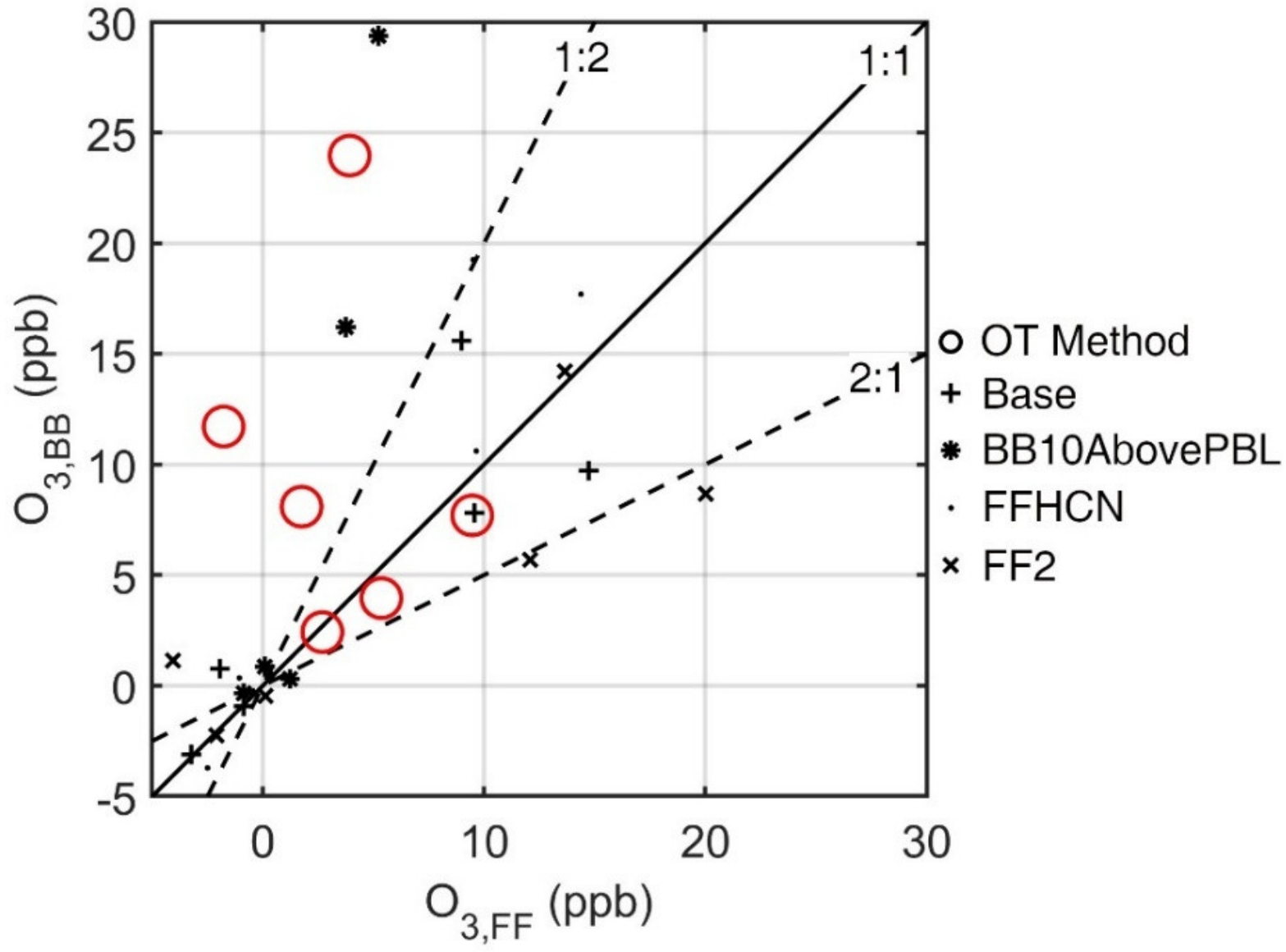

**Fig. S7. $O_3$ contributions from biomass burning (BB) or fossil fuel (FF) sources derived from the observation-based tracer approach of Bourgeois, et al. [3] (the OT method) with observational and simulated inputs.** Circles represent estimates derived from ATom observations, while other markers correspond to results using inputs from various model simulation scenarios. Base represents the standard model setup with default emission. BB10abovePBL scaled BB emissions by a factor of 10 and released them entirely above the planetary boundary layer (PBL) to account for uncertainty in BB emission. FF2 doubled the anthropogenic emissions to account for uncertainty in FF emission amount. FFHCN has HCN added to anthropogenic emissions to account for uncertainty in HCN emission ratios caused by possible emission of HCN from FF sources. See Materials and Methods for more details.

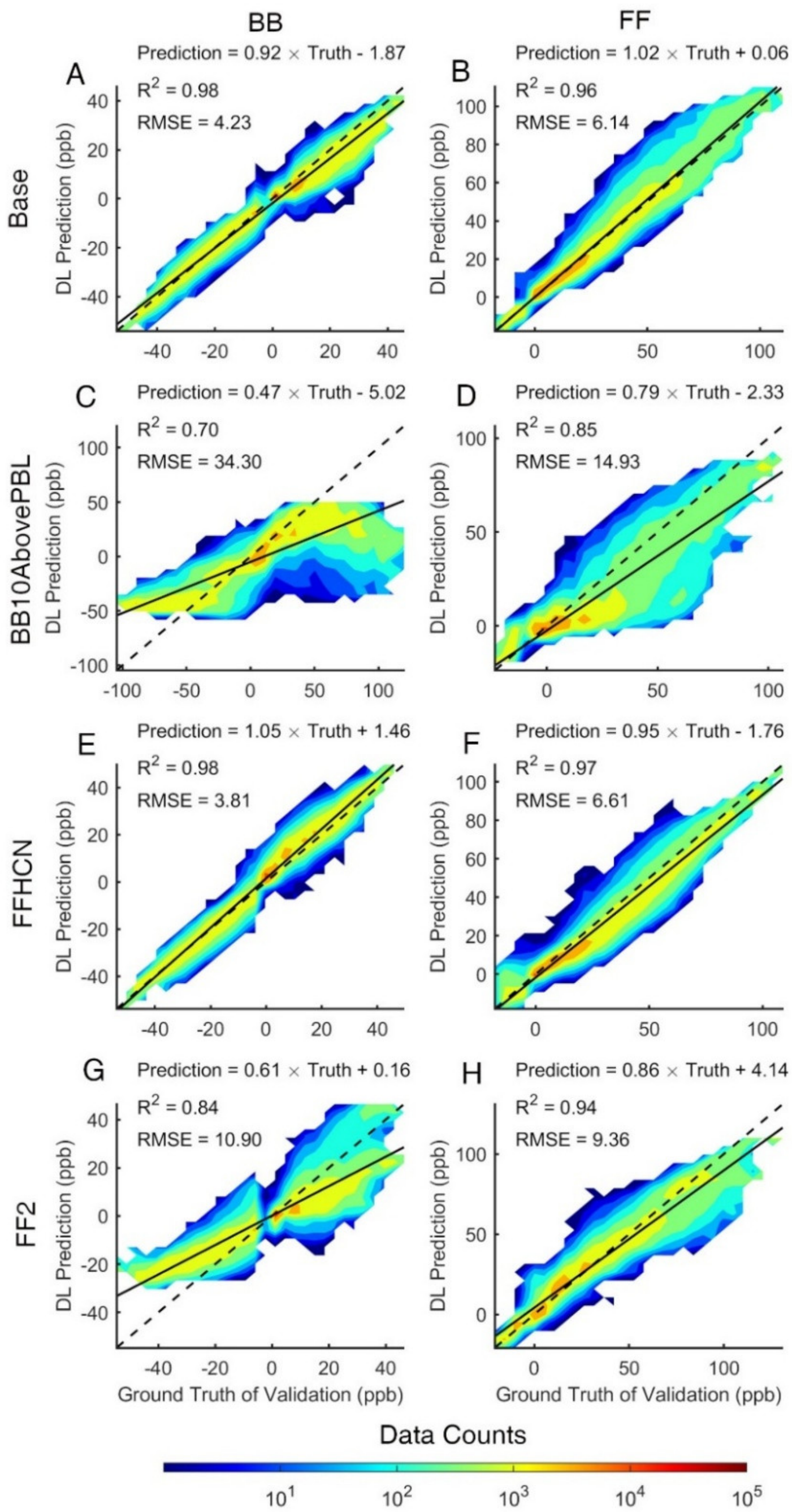

**Fig. S8. Results of cross-validation in an out-of-experiment style.** The DL model is validated with data from a certain simulation scenario (Base, BB10abovePBL, FFHCN, FF2, from top to bottom, see Materials and Methods) but trained without that scenario. The left and right column list the performance in predicting $O_3$ contributed by BB and FF emissions, respectively.

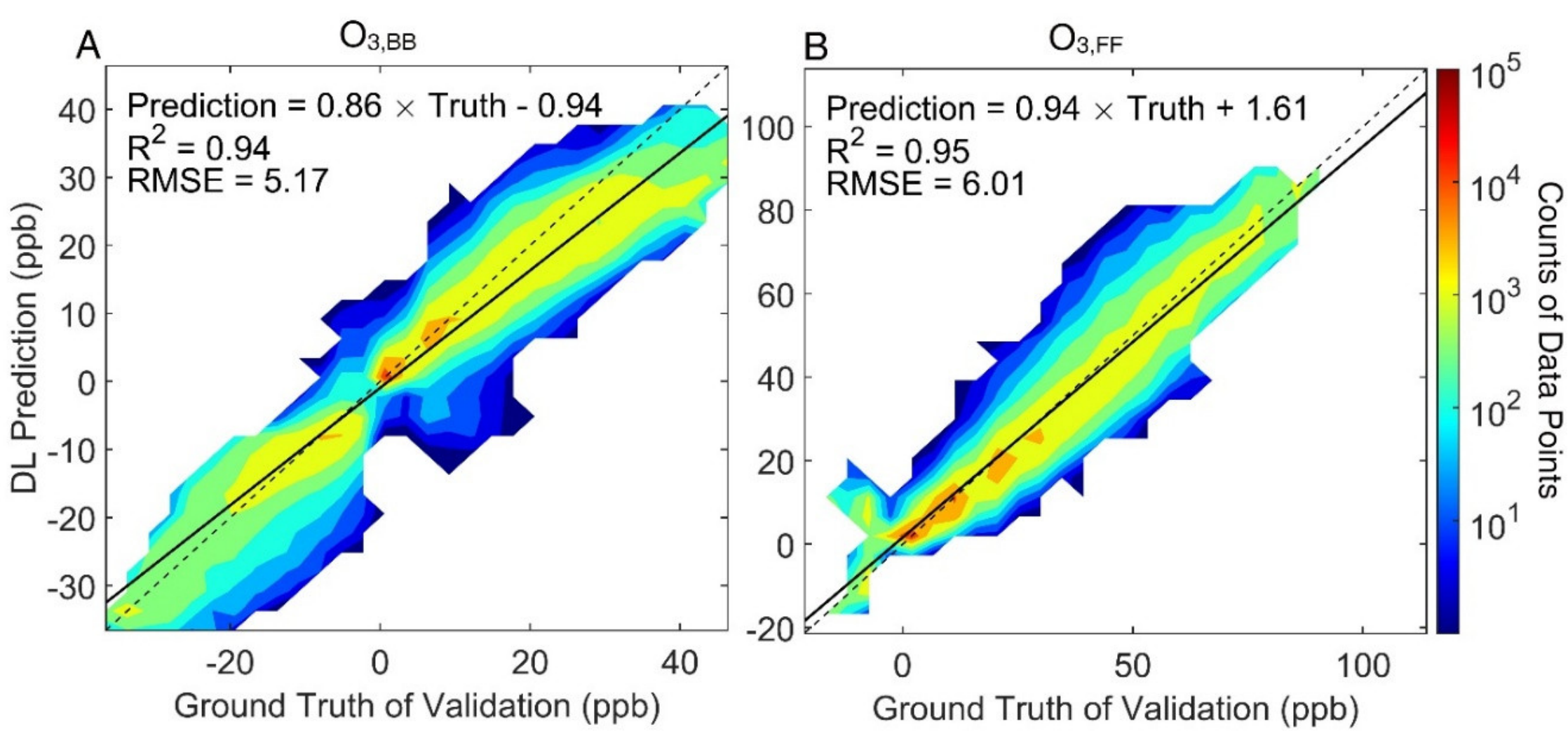

**Fig. S9. Validation of the deep learning (DL) model using data from an extended period.** The DL model was evaluated with data from an extended Base scenario simulation covering June-December 2018. The left and right column list the performance in predicting $O_3$ contributed by BB and FF emissions, respectively. The model had been trained using four simulation scenarios spanning June 2016-May 2018 as stated in the Methods and Materials.

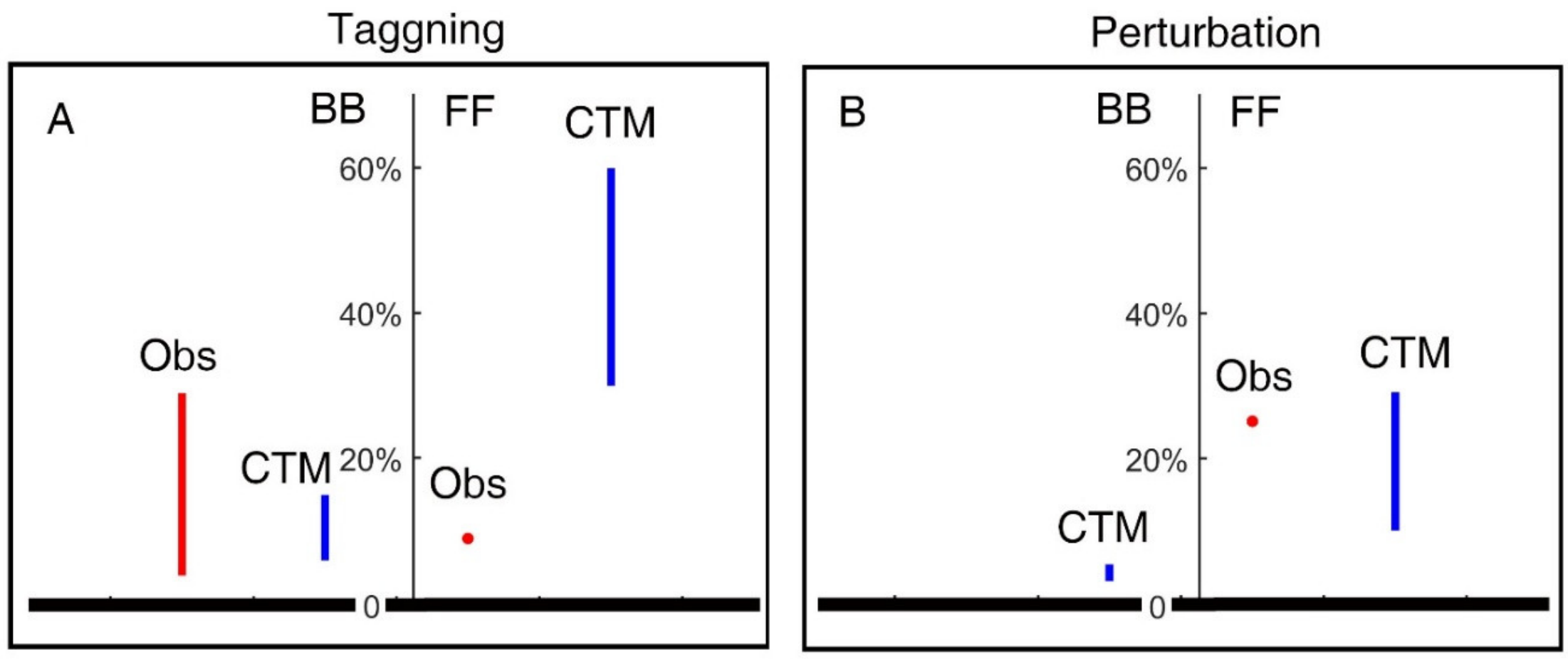

**Fig. S10. Summary of biomass burning (BB) and fossil fuels (FF) contributions to global tropospheric $O_3$ from previous studies described in *SI Appendix,* Sect. S1.** Comparisons are made between observation-based methods (Obs) and CTM-based approaches (CTM), categorized according to the tagging (left panel) and perturbation frameworks (right panel).

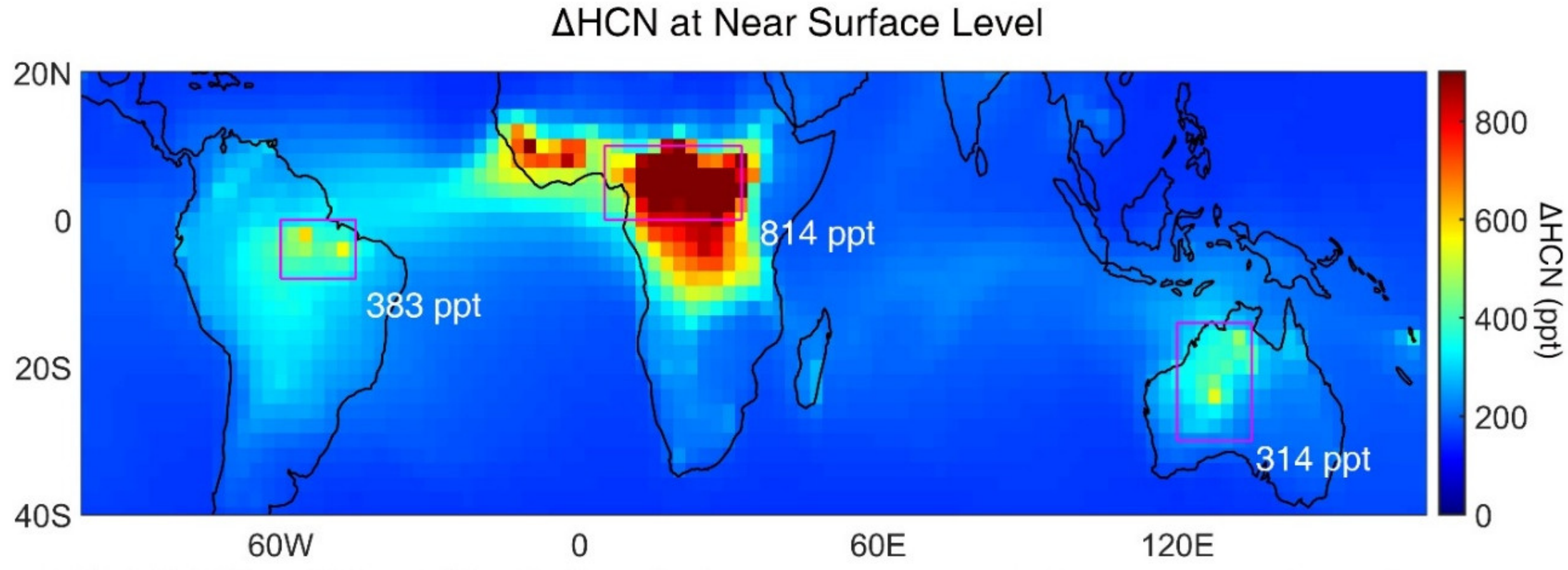

**Fig. S11. Near surface excess hydrogen cyanide mixing ratio (ΔHCN).** Simulated ΔHCN from Base Scenario was averaged over December 2017-January 2018. Grey text indicates the ΔHCN values averaged over the potential source regions (South America, Africa, and Australia, left to right), outlined by magenta rectangles.

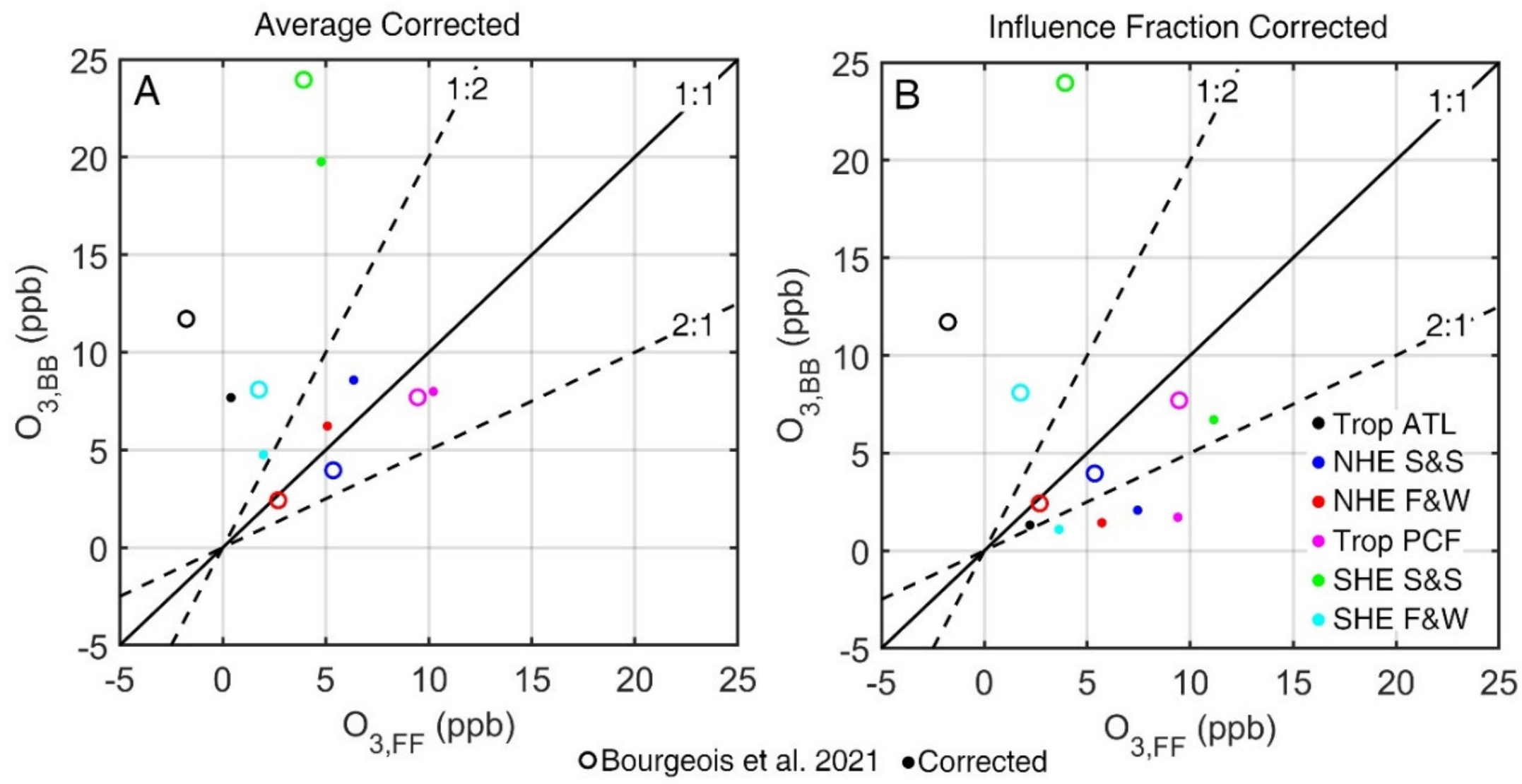

**Fig. S12. Change of the averaged $O_3$ contribution from biomass burning (BB) and fossil fuel (FF) after correcting potential issues in the observation-based tracer approach of Bourgeois, et al. [3] (the OT method).** (**A**) Correction for biases introduced by an invalid averaging process. (**B**) Correction for misinterpretation of the influence fraction. Definition of the regions and seasons can be found in the caption of Fig. 4. Detailed explanations of these corrections are provided in *SI Appendix,* Sect. S2.

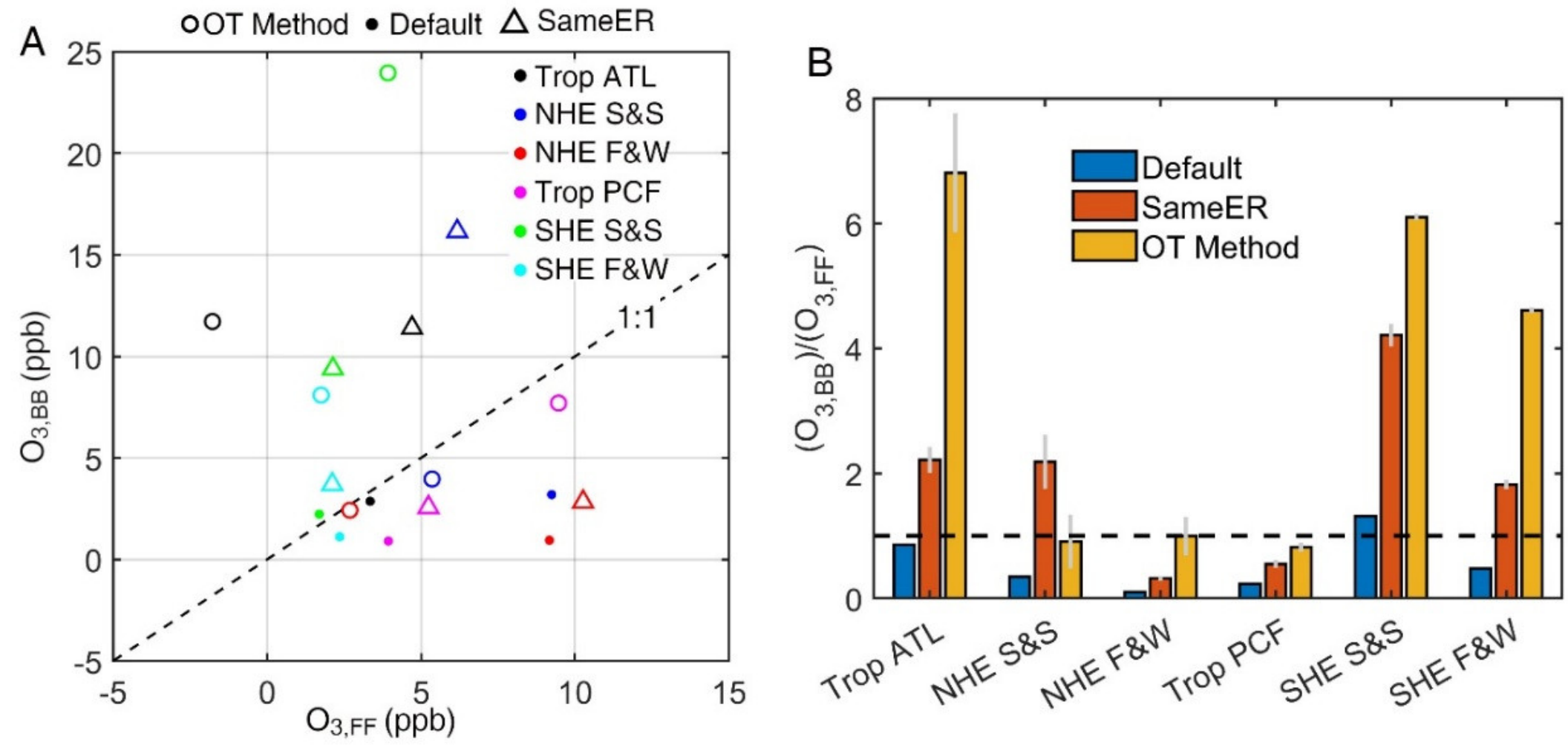

**Fig. S13. Changes in $O_3$ attribution when the deep learning (DL) model is trained with the SameER scenario.** SameER is an additional scenario in which CTM simulations were conducted by emitting NOx as a fixed ratio of CO for both BB and FF sources (see *SI Appendix,* Sect. S2). (**A**) Regional and global contributions of FF and BB sources to remote tropospheric $O_3$ in different seasons. Similar to Fig. 4A except the predictions from DL model trained with SameER are plotted as triangles. Our results from the default DL model (filled circles) and results from the OT method (unfilled circles) [3] are included for reference. (**B**) Ratio of $O_{3,BB}$ to $O_{3,FF}$ for different regions. For the OT method, the absolute value of the negative $O_{3,FF}$ in the Tropical Atlantic was used to calculate the ratio. Uncertainty ranges for SameER results reflect variation in emission ratios, while those for the OT method reflect different tracer pairs. See the caption of Fig. 4 for definition of regions.

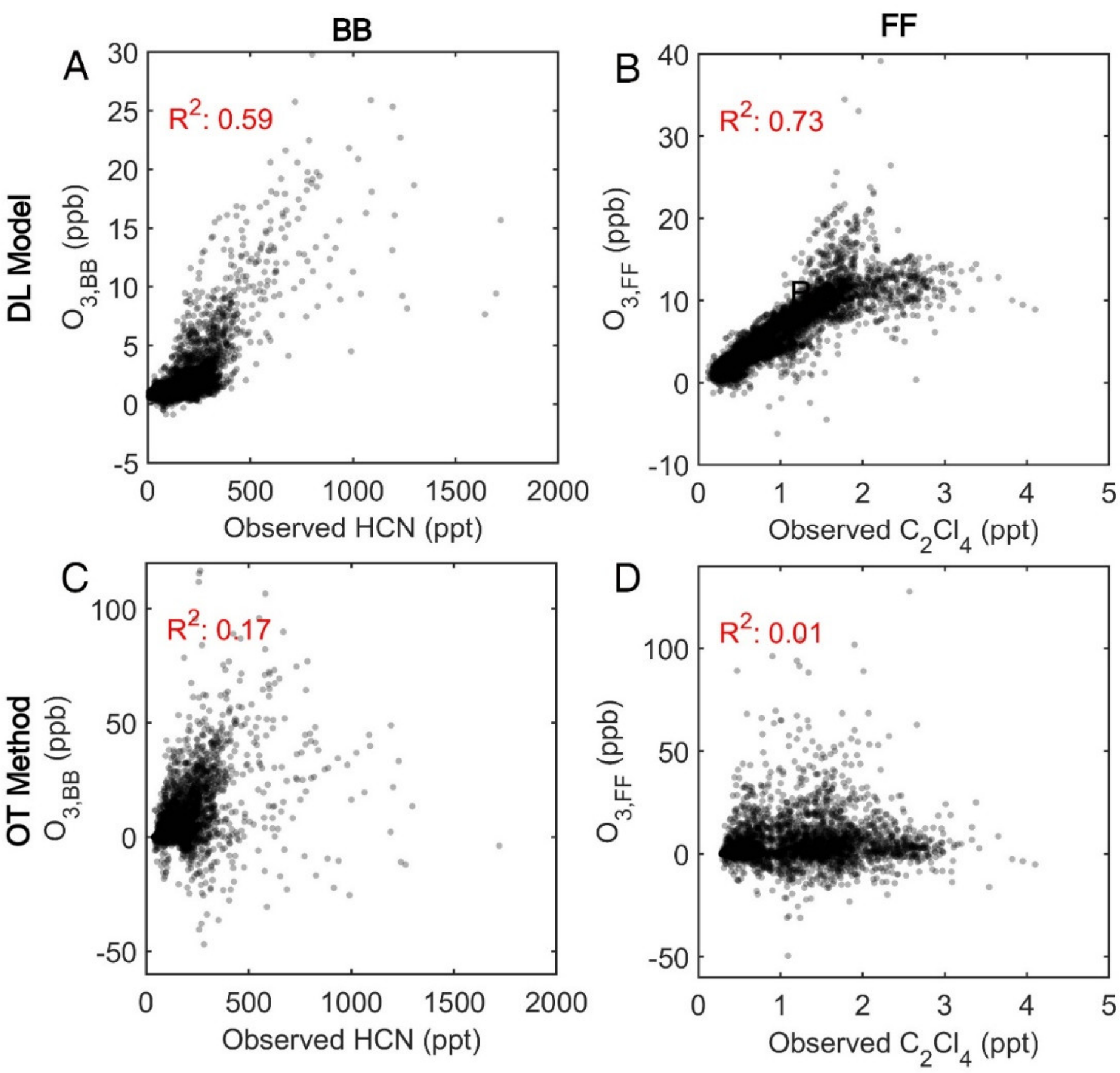

**Fig. S14. Correlation between the $O_3$ contribution from biomass burning (BB) and fossil fuel (FF) with their corresponding tracers.** (**A** and **B**) $O_3$ mixing ratios attributed to BB (A) or FF (B) sources by the DL model plotted against observed HCN (BB) or $C_2Cl_4$ (FF) mixing ratios for each ATom sample. (**C** and **D**) Same as (A, B), but with $O_3$ attributed using the OT method [3].

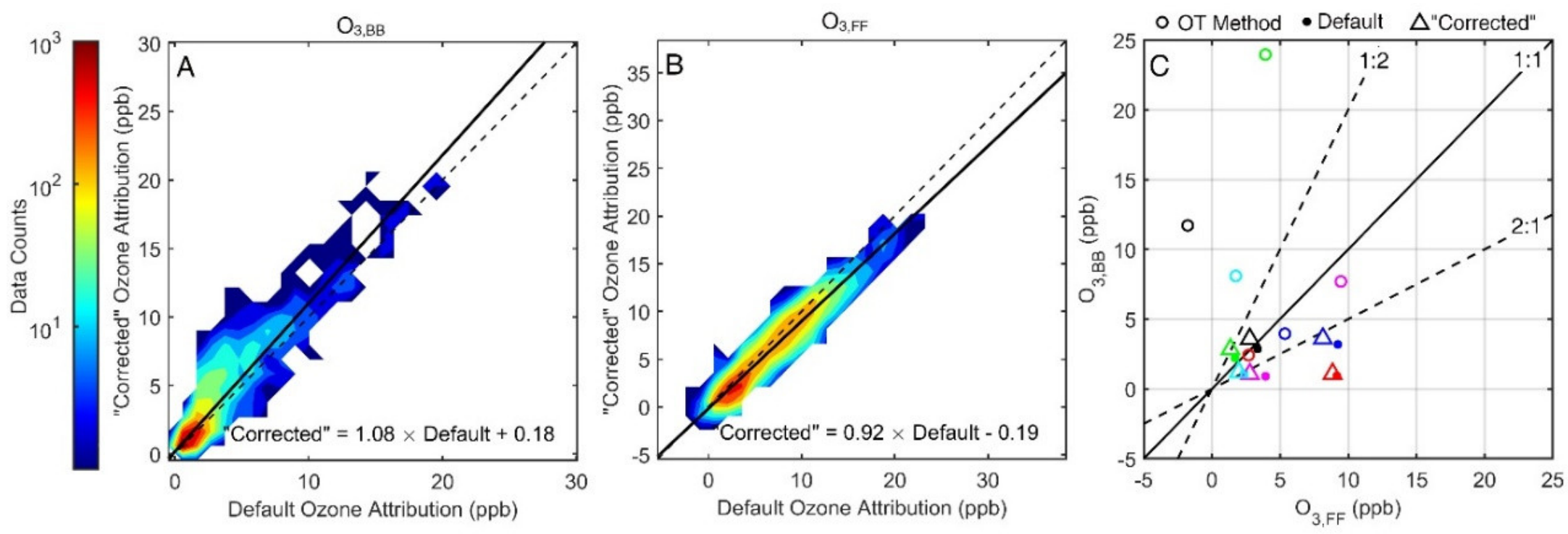

**Fig. S15. Sensitivity of deep learning (DL) model predictions to changes in emission ratios of source-specific tracers.** Default results are from the DL model trained as described in Materials and Methods. "Corrected" results are from a DL model trained on a dataset where HCN and $C_2Cl_4$ emission ratios were scaled by 0.8 and 1.2, respectively; both models use the same ATom observations as input. (**A** and **B**) Scatterplots comparing $O_3$ contributed by BB (A) or FF (B) between the default and "corrected" configurations. (**C**) Similar to Fig. 4A, but including results from the "corrected" configuration. Detailed explanations of these results are provided in *SI Appendix,* Sect. S4.

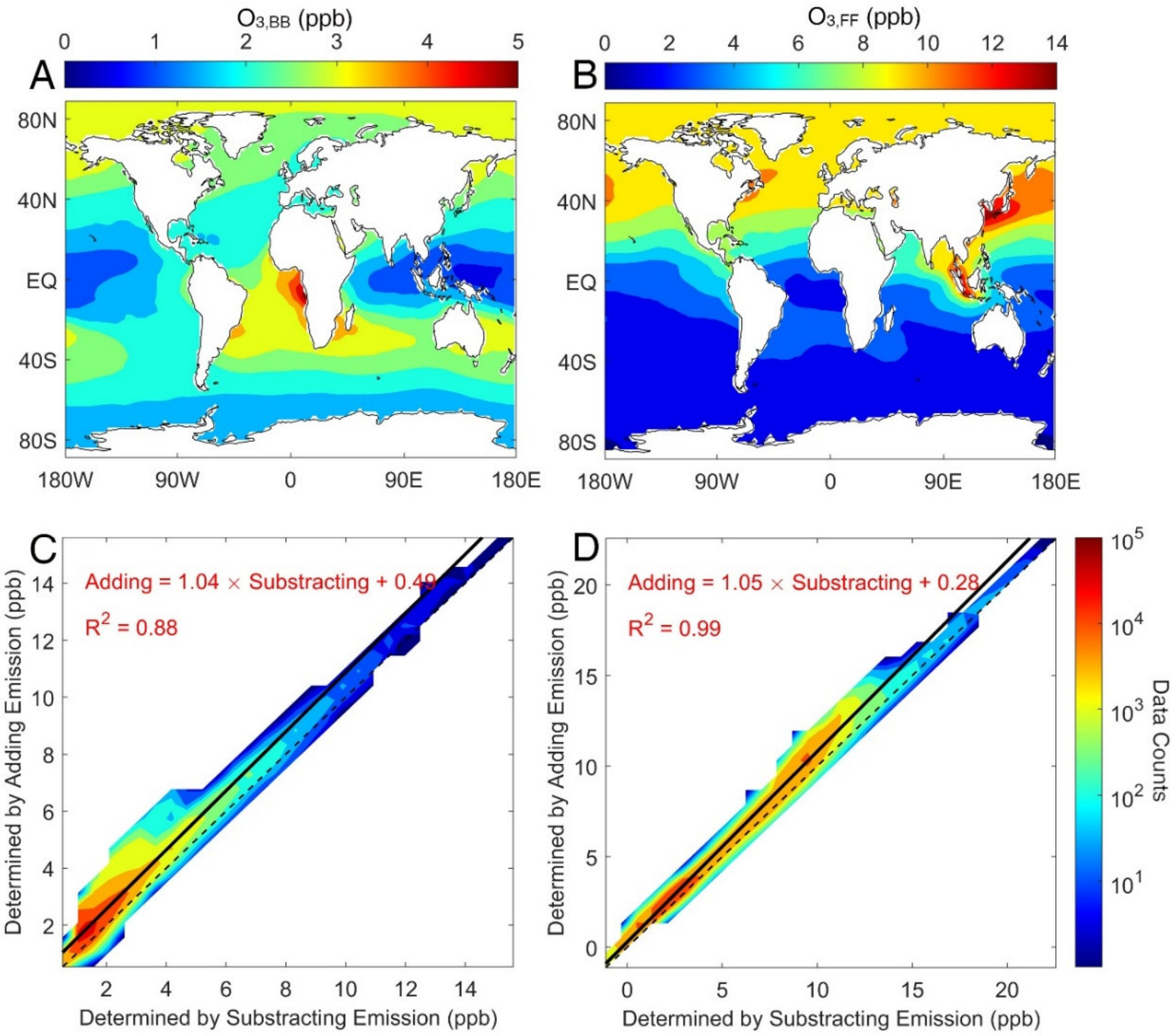

**Fig. S16. Influence of calculation approaches on $O_3$ contributions simulated by chemical transport models.** (**A** and **B**) Similar to Fig. S6A and S4B except the $O_3$ contributions are calculated using the "adding approach" instead of the "subtracting approach" (*SI Appendix,* Sect. S5). (**C** and **D**) Scatterplots comparing data points from Fig. S6A (Fig. S6B) with those in Panels A (B), illustrating the differences between the two calculation methods.

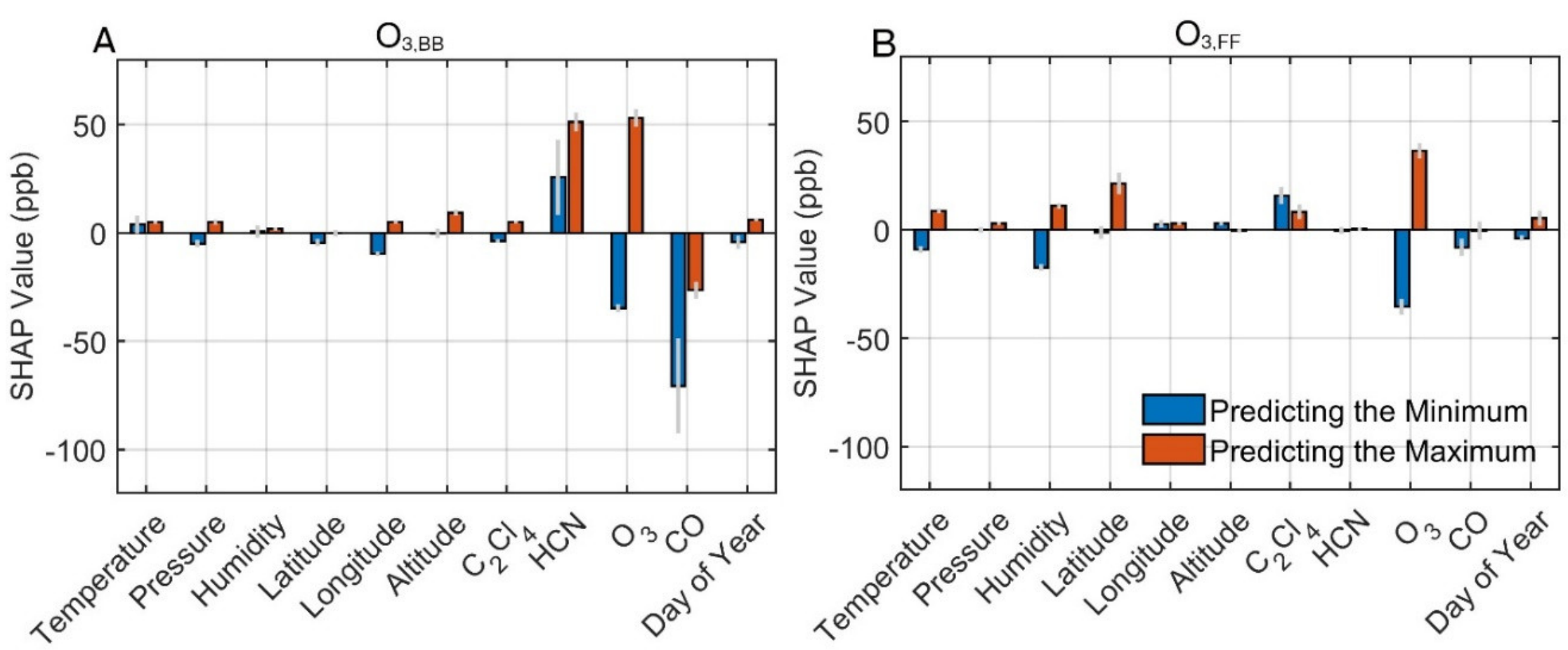

**Fig. S17. SHAP value maps for samples with maximum and minimum $O_3$ contributions from biomass burning (BB) and fossil fuel (FF) sources.** (**A**) and (**B**) show SHAP values for predictions corresponding to the maximum (orange) and minimum (blue) $O_3$ contributions from BB and FF sources, respectively. SHAP values were computed using a randomly subsampled dataset of 1000 samples from the training set. Gray lines indicate the uncertainty from subsampling, estimated by repeating the process 20 times. Detailed explanations of these results are provided in *SI Appendix,* Sect. S6.

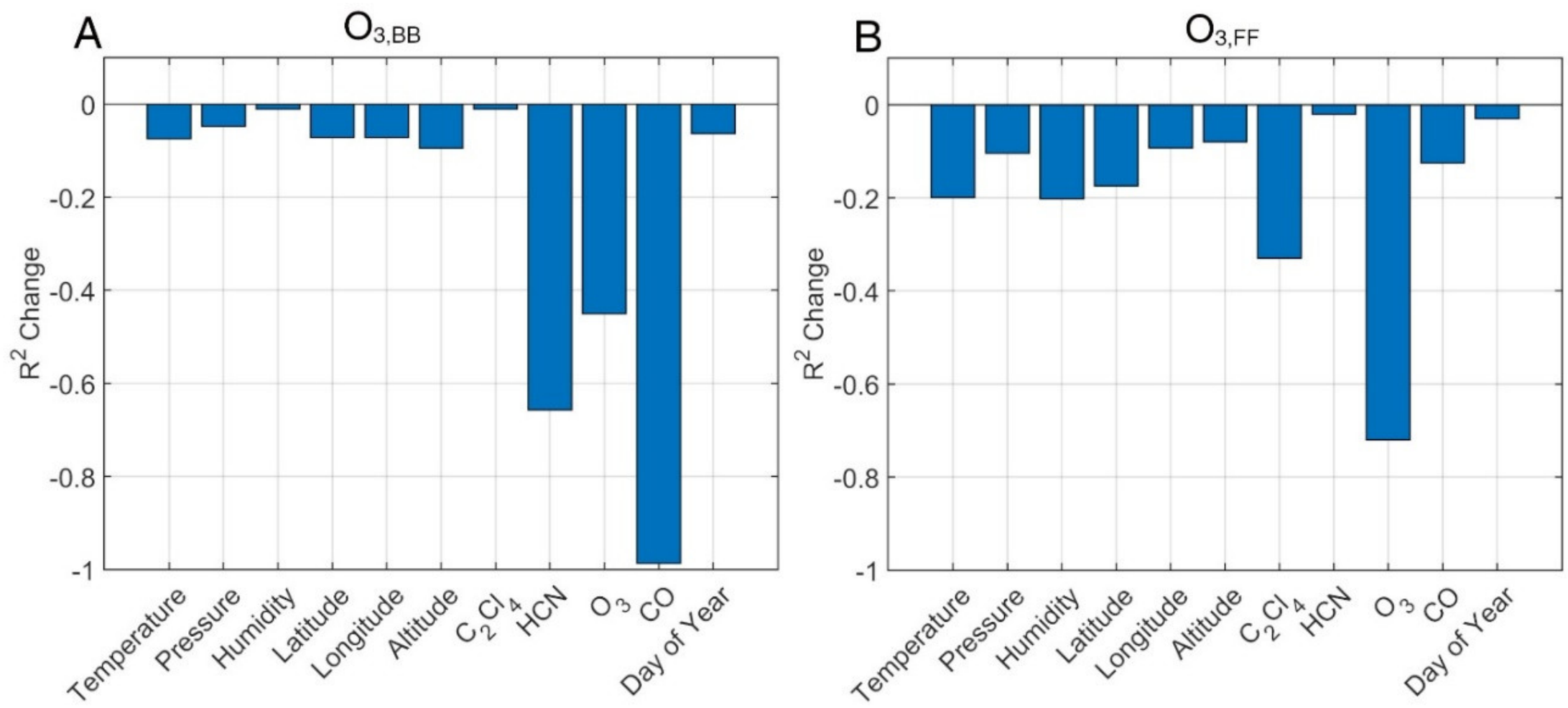

**Fig. S18. Partial dependence plots showing the importance of predictor variables for the DL model in estimating $O_3$ contributions from biomass burning (BB) and fossil fuel (FF) sources.** (**A**) and (**B**) show results for $O_3$ contributions from BB and FF sources, respectively. Each bar represents the change in coefficient of determination ($R^2$) after randomly permuting one predictor variable while keeping all others unpermuted. The resulting reduction in $R^2$ indicates the importance of that variable for model predictions. $R^2$ values are computed using the entire training dataset. Detailed explanations of these results are provided in *SI Appendix,* Sect. S6.

## SI References